\documentclass[11pt,a4paper]{article}

\usepackage{geometry} 

\usepackage{tikz-feynman}
\usepackage{xcolor}
\usepackage{float}
\usepackage{diagbox}
\usepackage{jheppub}
\usepackage{graphicx}
\usepackage{amsmath}
\usepackage{amssymb}
\usepackage{amsthm}
\usepackage{mathtools}
\usepackage{mathrsfs}

\usepackage{tikz}
\usepackage{subfigure,tikz}
\usetikzlibrary{backgrounds,automata}

\usepackage{subcaption}

\usepackage[T1]{fontenc} 
\usepackage{multirow}
\usepackage{booktabs}
\usepackage{natbib}
\usepackage{hyperref}
\usepackage{ulem}

\usepackage[title]{appendix}

\newcommand{\bea}{\begin{eqnarray}}
	\newcommand{\eea}{\end{eqnarray}}
\newcommand{\bean}{\begin{eqnarray*}}
	\newcommand{\eean}{\end{eqnarray*}}

\def\D #1{\dot{#1}}

\def\d{{\rm d}}

\def\a{{\alpha}}

\def\b{{\beta}}

\def\c{{\gamma}}

\def\d{\partial}

\allowdisplaybreaks

\def\Label#1{\label{#1}%
	\smash{\hbox to0pt{\raise1ex\hbox{\tiny[#1]}\hss}}}

\newcommand{\parall}[2]{{#1}\ /\kern -0.8em / \  {#2}}

\newcommand{\red}[1]{\textcolor{red}{{}#1}}

\title{\boldmath Generating Function of Loop Reduction by Baikov Representation}

\author[a,b,c]{Chang Hu,}

\affiliation[a]{College of Physics Science and Technology, Hebei University, Baoding 071002, China}

\affiliation[b]{Hebei Key Laboratory of High-precision Computation and Application of Quantum Field Theory, Baoding,
071002, China
}
\affiliation[c]{Hebei Research Center of the Basic Discipline for Computational Physics, Baoding, 071002, China}

\author[d,e,f]{Wen-Di Li*,}

\affiliation[d]{School of Fundamental Physics and Mathematical Sciences, Hangzhou Institute for Advanced Study, UCAS, Hangzhou 310024, China}

\affiliation[e]{University of Chinese Academy of Sciences, Beijing 100049, China}

\affiliation[f]{Institute of Theoretical Physics, Chinese Academy of Sciences, Beijing 100190, China}

\author[g]{Xiang Li.}

\affiliation[g]{School of Physics, Peking University, Beijing 100871, China}

\emailAdd{isiahalbert@126.com} \emailAdd{liwendi23@mails.ucas.ac.cn} 
\emailAdd{lix-PHY@pku.edu.cn}

\abstract{The generating function method was recently introduced for the reduction of loop integrals \cite{Feng:2022hyg}. In this work, we integrate this method with the residue calculus in the Baikov representation. We demonstrate that this combined approach leads to simpler computations and more compact expressions compared to the momentum-space representation \cite{Feng:2024qsa}. Furthermore, the method's applicability extends beyond standard one-loop integrals with quadratic propagators; we provide explicit examples for integrals with linear propagators and for several multi-loop cases. These results suggest that suitable representations, such as the Baikov representation and generating functions, make the underlying analytic structure of Feynman integrals more apparent, facilitating further analysis with tools from analytic function theory.}

\renewcommand{\thefootnote}{\fnsymbol{footnote}}
\footnotetext[1]{Corresponding author}

\usepackage{amsfonts}
\begin{document}

\maketitle

\section{Introduction}\label{sec: 1}

Scattering amplitudes hold central importance in quantum field theory as they bridge the gap between theoretical predictions and experimental observations. With the successful operation of the Large Hadron Collider (LHC) \cite{Azzi:2019yne,Cepeda:2019klc} and proposals for next-generation collider concepts \cite{ILC:2013jhg,Behnke:2013xla,Bambade:2019fyw,CEPCStudyGroup:2018ghi,CEPCStudyGroup:2018rmc,TLEPDesignStudyWorkingGroup:2013myl,FCC:2018byv,FCC:2018evy}, extending perturbative calculations of scattering processes to higher orders (such as next-to-next-to-leading order) requires computing multi-loop scattering amplitudes. Utilizing Lorentz symmetry, these amplitudes can be expressed as linear combinations of scalar Feynman integrals (FIs). The computation of scalar Feynman integrals constitutes a major challenge in cutting-edge research.

Studies show that a family of Feynman integrals forms a finite-dimensional linear space with basis elements called master integrals (MIs). Thus, current mainstream methods for computing scalar Feynman integrals involve two separate tasks: The first is Feynman integral reduction, aiming to express Feynman integrals as linear combinations of basis MIs \cite{Chetyrkin:1981qh, Gehrmann:1999as, Laporta:2000dsw, Baikov:2005nv, Baikov:2007zza, Gluza:2010ws,Schabinger:2011dz, Larsen:2015ped, Wu:2023upw, vonManteuffel:2014ixa, Peraro:2016wsq,  Peraro:2019svx, Klappert:2019emp, Klappert:2020aqs,  Belitsky:2023qho, Liu:2018dmc, Guan:2019bcx, Guan:2024byi, Smirnov:2020quc, Anastasiou:2004vj, Smirnov:2008iw,Smirnov:2023yhb,Maierhofer:2017gsa, Klappert:2020nbg, Lee:2012cn, Lee:2013mka, Studerus:2009ye,vonManteuffel:2012np,   Mastrolia:2018uzb, Frellesvig:2019kgj,  Frellesvig:2020qot,Weinzierl:2020xyy, Wang:2019mnn, Boehm:2020ijp, Basat:2021xnn,Feng:2021enk,Hu:2021nia,Feng:2022uqp,Feng:2022iuc,Feng:2022rfz}; the second is computing these MIs \cite{Hepp:1966eg,Roth:1996pd,Binoth:2000ps,Heinrich:2008si,Smirnov:2008py,Smirnov:2021rhf,Carter:2010hi,Borowka:2017idc,Borowka:2018goh,Bergere:1973fq,Boos:1990rg,Smirnov:1999gc,Tausk:1999vh,Czakon:2005rk,Smirnov:2009up,Gluza:2007rt,Jantzen:2012mw,Lee:2009dh,Kotikov:1990kg,Kotikov:1991pm,Remiddi:1997ny,Gehrmann:1999as,Argeri:2007up,Muller-Stach:2012tgj,Henn:2013pwa,Henn:2014qga,Moriello:2019yhu,Hidding:2020ytt,Catani:2008xa,Bierenbaum:2010cy,Bierenbaum:2012th,Tomboulis:2017rvd,Runkel:2019yrs,Capatti:2019ypt,Aguilera-Verdugo:2020set,Panzer:2015ida,Panzer:2014caa,Hidding:2022ycg,Song:2021vru,Zeng:2023jek,Liu:2017jxz,Liu:2020kpc,Liu:2021wks,Liu:2022mfb,Liu:2022tji,Liu:2022chg}. Notably, based on the auxiliary mass flow method \cite{Liu:2017jxz,Liu:2020kpc, Liu:2021wks,Liu:2022chg,Liu:2022tji,Liu:2022mfb}, any given Feynman integral can automatically be calculated to high precision once its reduction is completed. However, in complex multi-loop processes, integral reduction remains a critical and highly challenging step.

Integration-by-parts (IBP) identities \cite{Chetyrkin:1981qh,Kosower:2018obg} combined with the Laporta algorithm \cite{Laporta:2000dsw} serve as the primary approach for quantitative reduction. Although strategies like finite field methods \cite{vonManteuffel:2014ixa, Peraro:2016wsq,  Peraro:2019svx, Klappert:2019emp, Klappert:2020aqs,  Belitsky:2023qho} (to avoid intermediate expression growth), syzygy equations \cite{Gluza:2010ws,Schabinger:2011dz,Larsen:2015ped,Bohm:2018bdy} and block-triangular form \cite{Liu:2018dmc,Guan:2019bcx,Guan:2024byi} (to reduce IBP system size) have been adopted, the reduction of Feynman integrals with high-degree denominators or irreducible scalar products (ISPs) still demands excessive time and computational resources. Other methods include well-known PV reduction \cite{Passarino:1978jh}, OPP method \cite{Ossola:2006us,Ossola:2007ax}, computational algebraic geometry methods \cite{Mastrolia:2011pr,Badger:2012dp} and unitarity cuts \cite{Bern:1994zx,Britto:2004nc,Britto:2005ha,Britto:2006sj,Anastasiou:2006jv,Britto:2006fc,Britto:2010um} . Several new methods have been proposed to avoid or partially bypass IBP reduction, but each faces inherent difficulties: In frameworks based on intersection theory \cite{Mastrolia:2018uzb, Frellesvig:2019kgj,  Frellesvig:2020qot,Weinzierl:2020xyy}, computing intersection numbers for multivariate problems remains challenging; methods using large spacetime \cite{Baikov:2005nv,Baikov:2007zza} or auxiliary mass \cite{Liu:2018dmc, Wang:2019mnn} expansions struggle to obtain high-order terms; approaches employing fixed-branch-integral representations \cite{Huang:2024nij}, while converting multi-loop integrals into one-loop-like integrals with integration of branch parameters to simplify IBP reduction, face difficulties in achieving high-precision integration.

In the various reduction methods mentioned above, a common feature is the presence of an iterative structure. In \cite{Feng:2022hyg}, the authors proposed using the generating function approach to handle these structures. The core idea of the generating function is to introduce one or several appropriate auxiliary parameters to encapsulate the recurrence relations of the reduction, thereby providing a unified representation of these iterative systems. In \cite{Hu:2023mgc,Hu:2025spa}, we addressed the reduction of one-loop tensor Feynman integrals using generating functions and derived the general formula, avoiding the complexity of iterative computations. In \cite{Feng:2024qsa}, we utilized IBP relations to propose a generating function method for calculating the reduction of one-loop higher-order poles and provided general formulas for reductions to Master Integrals with the  top and sub top sectors. Specifically, for a one-loop integral with $n$ propagators and higher-order poles, we construct the following generating function:
\begin{small}
    \begin{equation}
    \begin{aligned}
        \int \frac{d^Dl}{D_1^{a_1} D_2^{a_2}\cdots D_n^{a_n}}\to \sum_{a_1,\cdots,a_n=1}^{\infty}\int d^Dl\frac{t_1^{a_1-1}t_2^{a_2-1}\cdots t_n^{a_n-1}}{D_1^{a_1}D_2^{a_2}\cdots D_n^{a_n}}=\int \frac{d^Dl}{(D_1-t_1)\cdots (D_n-t_n)}.
    \end{aligned}
    \label{eq:GF01}
\end{equation}
\end{small}
Then the corresponding reduction coefficients turn to:
\begin{equation}
    C_{i}(a_1,a_2,\cdots,a_n)\to \sum_{a_1,\cdots,a_n=1}^\infty C_{i}(a_1,a_2,\cdots,a_n)t_1^{a_1-1}t_2^{a_2-1}\cdots t_n^{a_n-1}\equiv GF_i(t_1,t_2,\cdots,t_n).
    \label{eq:GF02}
\end{equation}

In \cite{Feng:2024qsa,Li:2024rvo}, the generating function based on IBP satisfies a system of partial differential equations. By exploiting the solvability of these equations, we transformed the original complex system of multivariable partial differential equations into a simpler ordinary differential equation, which only involves a single variable. As a result, the solution takes the form of a single-variable integral. However, in the computation of reductions to the sub top sector, we did not use this integral form. Instead, by analyzing the analytic structure of the matrices appearing in the partial differential equation system, we derived a relatively simple expression. See Equation (6.35) in Section 6.2 of \cite{Feng:2024qsa}, which contains only one hypergeometric function, with the remaining terms being rational functions. On the other hand, the integral form of the generating function directly written from the equations is significantly more complex. See Equation (3.53) in Section 3.2 of \cite{Li:2024rvo}, where the expression is an infinite series, with each term containing an Appell function. Both forms yield correct results, which at least indicates that not all the analytic information has been fully utilized.

In fact, the results from these two papers show a strong similarity between the generating function for reductions to the top sector and the Baikov polynomial in the Baikov representation. This similarity is not difficult to understand, as the generating function is essentially a series expansion of the reduction coefficients, while in the Baikov representation, the reduction coefficients correspond to the residues of the higher-order poles at $z_1 = z_2 = \cdots = z_n = 0$. Therefore, they are equivalent to the Taylor series expansion coefficients of the Baikov polynomial at the corresponding pole orders.

This naturally leads us to consider that, similarly based on IBP, there should be a simpler way to compute generating functions in the Baikov representation \cite{Baikov:1996rk,Baikov:2000jg,Baikov:1998ax,Smirnov:2003kc}. Compared to the momentum-space representation  of Feynman integrals, the Baikov representation is unified for both one-loop and multi-loop cases, and it provides a unified treatment of propagators, even when they are non-standard quadratic forms in the loop momentum. Therefore, we believe that using the IBP form in the Baikov representation to compute generating functions may facilitate extensions to higher-loop cases or propagators with non-standard quadratic forms, potentially leading to new results. This is the core idea of this paper.

In section \ref{sec: 2}, we briefly review the Baikov's approach to the IBP reduction, to set the stage for later developments. Section \ref{sec: 3} introduces the construction of generating functions within the Baikov representation. In section \ref{sec: 4}, we formulate a general framework for computing generating functions using this representation, providing a systematic approach to the problem. Sections \ref{sec: 5} and \ref{sec: 6} present explicit examples at one-loop and higher-loop levels, respectively, illustrating the applicability of the method in concrete settings. Section \ref{sec: 7} is devoted to explaining how to extract reduction coefficients from the generating functions, highlighting several subtle but important aspects that require attention. Finally, in section \ref{sec: 8}, we offer a summary and discussion of this work.

\section{Review}\label{sec: 2}

In this section, we review not only the Baikov representation itself, but also how it can be used to compute reduction coefficients via residue evaluation. This provides a foundation for the generating function approach developed in the following sections.

\subsection{Baikov representation}\label{sec: 2.1}

The Baikov representation reformulates Feynman integrals by introducing variables that are directly tied to the scalar products of momentum, making it particularly useful for complex multi-loop integrals. Here’s a step-by-step breakdown of its formulation and utility.

A typical $L$-loop Feynman integral with $N$ propagators is written as:
\begin{equation}
I(a_1,a_2,\cdots,a_n)=\int \prod_{i=1}^L d^D l_i \frac{1}{D_1^{a_1} D_2^{a_2} \cdots D_N^{a_N}},
\label{eq:tradition form}
\end{equation}
where:
\begin{itemize}
    \item $l_i$ are the loop momenta,
    \item The denominators corresponding to the propagators are given by  
\begin{equation}
    D_i = \sum_{\alpha\leq \beta} A_{i}^{\alpha\beta} (l_{\alpha} \cdot l_{\beta}) + \sum_{\alpha,\beta} B_{i}^{\alpha\beta} (l_{\alpha} \cdot q_{\beta}) + M^2_i,
    \label{eq:gen propagator}
\end{equation} 
where $q_{\beta}$ are the external momentum. This form of $D_i$ represents a very general structure for propagators, allowing for broad applications beyond standard quadratic forms.

\item The $a_i$ are the propagator powers, which are allowed to be positive integers, negative integers, or zero.

\item Here, we emphasize that the number of propagators $N$ is chosen as $L(L+1)/2+LE$, where $E$ is the number of independent external momentum, ensuring that all scalar products involving loop momenta can be expressed as a linear combination of the propagators. For cases with fewer propagators, the corresponding powers can simply be set to $a_i = 0$. For cases with more propagators, they can be reduced to a form with $N$ propagators using factorization or Mellin-Barnes decomposition.

\end{itemize}

In the Baikov representation, the integration variables are transformed from  $l_i$(loop momenta) to new variables $z_i\equiv D_i$ represent the propagators directly. The measure becomes: 
\begin{equation}
\prod_{i=1}^L d^D l_i \rightarrow \prod_{i=1}^N d z_i P(\vec{z})^{(D-L-E-1) / 2},
\end{equation}
where $P(\vec{z})$ is the Baikov polynomial, which is defined as the Gram determinant 
\begin{equation}
    \text{det} G(l_1,l_2,\cdots,l_L,q_1,q_2,\cdots,q_E),
\end{equation}
where all scalar products involving loop momenta are expressed as linear combinations of Baikov variables(i.e., propagators).  For example, for a one-loop triangle with three propagators carrying the same mass, we have the following expressions for the propagator denominators:
\begin{equation}
     z_1=l^2-m^2,\ \ z_2=(l-q_1)^2-m^2,\ \ z_3=(l-q_2)^2-m^2.
\end{equation}
We can then express the three scalar products involving the loop momentum $l$ as a linear combination of $z_1$, $z_2$, and $z_3$:
\begin{equation}
    l^2=z_1+m^2 ,\ \ \ 
    l\cdot q_1 = \frac{z_2-z_1-q_1^2}{2} ,\ \ \ 
    l\cdot q_2 =\frac{z_3-z_1-q_2^2}{2} .
\end{equation}
Then the Baikov polynomial is given by
\begin{align}
      P(z_1,z_2,z_3) &=  \text{det}G(l,q_1,q_2) =
\begin{vmatrix} 
  l^2    &   l\cdot q_1   &   l\cdot q_2   \\
  l\cdot q_1    &   q_1^2  &   q_1\cdot q_2 \\
  l\cdot q_2      &   q_1\cdot q_2    &   q_2^2
\end{vmatrix}  = 
\begin{vmatrix} 
  z_1+m^2    &       \frac{z_2-z_1-q_1^2}{2}    &    \frac{z_3-z_1-q_2^2}{2} \\
  \frac{z_2-z_1-q_1^2}{2}      &       q_1^2  & q_1\cdot q_2      \\
  \frac{z_3-z_1-q_2^2}{2}     &       q_1\cdot q_2        &   q_2^2
\end{vmatrix}  \nonumber\\
&=  \frac{1}{4}\Big[
  -(q_1-q_2)^2z_1^2 -q_2^2z_2^2 -q_1^2z_3^2 +2(q_1^2-q_1\cdot q_2)z_1z_3
  +2(q_2^2-q_1\cdot q_2)z_1z_2 \nonumber\\
& +2q_1\cdot q_2z_2z_3  +2q_1\cdot q_2(q_1-q_2)^2z_1  
  +2q_2^2(q_1^2-q_1\cdot q_2)z_2  +2q_1^2(q_2^2-q_1\cdot q_2)z_3  \nonumber\\
& +4m^2(q_1^2q_2^2-(q_1\cdot q_2)^2)  -q_1^2q_2^2(q_1-q_2)^2
\Big].\label{eq:example of Baikov Poly }
\end{align}
After defining the Baikov polynomial, the original Feynman integral \eqref{eq:tradition form} in the Baikov representation can be written as:

\begin{equation}
  I(a_1,...,a_{N}) = 
  \int_{\Gamma} d^{N}z [P(\vec{z})]^{\frac{D-L-E-1}{2}}\frac{1}{z_1^{a_1} z_2^{a_2}...z_N^{a_N}}.
  \label{eq:baikov form}
\end{equation}
The integration region $\Gamma$ is constrained to the physical domain.

\subsection{Reduction Coefficient}\label{sec: 2.2}

The Baikov representation is primarily valued for its clarity in analyzing IBP structures and identifying master integrals. 
An Feynman integral can be reduced to a linear combination of some master integral as:
\begin{equation}
    I(a_1,a_2,\cdots ,a_N) = \sum_i C_i(a_1,a_2,\cdots ,a_N) I^i(a_1^i,a_2^i,\cdots ,a_N^i).
    \label{eq:Reduce to master integrals}
\end{equation}
In the Baikov representation, the above formula becomes: 
\begin{footnotesize}
    \begin{equation}
    \int_{\Gamma} \mathrm{d}^{N}z [P(\vec{z})]^{\frac{D-L-E-1}{2}}\frac{1}{z_1^{a_1} z_2^{a_2}...z_N^{a_N}} = \sum_i C_i(a_1,\cdots,a_N) \int_{\Gamma} d^{N}z [P(\vec{z})]^{\frac{D-L-E-1}{2}}\frac{1}{z_1^{a_1^i} z_2^{a_2^i}\cdots z_N^{a_N^i}}.
    \label{eq:Baikov red 01}
\end{equation}
\end{footnotesize}

We assume that there are $M$ non-zero master integrals on the right-hand side of \eqref{eq:Baikov red 01}. By taking residues at $z_i = 0$ for different subsets of the variables $\{z_1,z_2,...,z_N\}$ on both sides, we can construct a system of linear equations for the reduction coefficients.\footnote{Here the “residue equations” are understood as a residue-based projection of the IBP identities in Baikov variables; total-derivative terms integrate to zero on the Baikov domain, which justifies taking residues on both sides.} Naively, this approach yields $2^N - 1$ equations. However, these equations are not always independent. For example, in cases where certain pole powers $a_i = 0$, some equations may reduce to trivial identities like $0 = 0$. Nevertheless, as long as the number of independent equations, $N_{\text{ind}}$, equals $M$, it is possible to construct a sufficient number of equations to fully determine the reduction coefficients.

However, if we later construct a generating function in the form of \eqref{eq:GF01} \eqref{eq:GF02}, then all the poles $a_i$ will be positive integers. In the case where the number of propagators is $N = L(L+1)/2 + LE$, all scalar products can be expressed as linear combinations of the propagators. In this scenario, the propagator exponents (or exponents of baikov variables $z_i$) in the corresponding master integrals will either be 0 or 1. 

In this case, the reduction coefficients are computed iteratively by taking residues at $z_i = 0$, starting with all $N$ values of $z_i=0$, then progressively reducing the number of $z_i=0$ to $N-1$, and continuing in this manner until only one $z_i = 0$ remains.

To be detailed, first we take the residues of $z_i=0$  of all $N$ Baikov variables. On the left-hand side of \eqref{eq:Baikov red 01}, taking the residue involves calculating the Taylor expansion coefficient of the denominator at the origin, which is
\begin{footnotesize}
    \begin{equation}
  \frac{ 1}{ (a_{1}-1)!(a_{2}-1)!...(a_N-1)! }
  \Big(\frac{\partial}{\partial z_{1}}\Big)^{a_{1}-1}
  \Big(\frac{\partial}{\partial z_{2}}\Big)^{a_{2}-1}...
  \Big(\frac{\partial}{\partial z_{N}}\Big)^{a_{N}-1}
  [P(\vec{z})]^{\frac{D-L-E-1}{2}}\Big|_{z_1,...,z_N=0}.
\end{equation}
\end{footnotesize}
On the right-hand side, the only remaining term is $a_1^i = a_2^i = \cdots = a_N^i = 1$. 
\begin{equation}
    C_N(a_1,a_2,\cdots,a_N)\cdot [P(\vec{z}=\vec{0})]^{\frac{D-L-E-1}{2}} .
\end{equation}
For other master integrals, the presence of Baikov variables with $a_j^i = 0$ leads to the absence of singularities, causing the residues to vanish. This process ultimately yields the reduction coefficient for the top sector.
\begin{footnotesize}
    \begin{equation}
    C_N(a_1,a_2,...,a_N) = 
  \frac{ [P(\vec{0})]^{-\frac{D-L-E-1}{2}} }{ (a_{1}-1)!...(a_N-1)! }
  \Big(\frac{\partial}{\partial z_{1}}\Big)^{a_{1}-1}
  ...
  \Big(\frac{\partial}{\partial z_{N}}\Big)^{a_{N}-1}
  [P(\vec{z})]^{\frac{D-L-E-1}{2}}\Big|_{z_1,...,z_N=0}.
\end{equation}
\end{footnotesize}
Second, we take the residues of  $N-1$ variables $\{z_1,...,z_{j-1},z_{j+1},...,z_N\}=\vec{0}$, and the equation transforms into
\begin{align}
      \int_{\Gamma} dz_j&
    \frac{1}{z_j^{a_j}} \frac{1 }{ (a_{1}-1)!...(a_{j-1}-1)!(a_{j+1}-1)!...(a_N-1)! }
  \Big(\frac{\partial}{\partial z_{1}}\Big)^{a_{1}-1}
  \Big(\frac{\partial}{\partial z_{2}}\Big)^{a_{2}-1}...
  \Big(\frac{\partial}{\partial z_{j-1}}\Big)^{a_{j-1}-1} \nonumber\\
 &\times \Big(\frac{\partial}{\partial z_{j+1}}\Big)^{a_{j+1}-1}...
  \Big(\frac{\partial}{\partial z_{N}}\Big)^{a_{N}-1}
  [P(\vec{z})]^{\frac{D-L-E-1}{2}}\Big|_{z_1,z_2,...,z_{j-1},z_{j+1},...,z_N=0} \nonumber\\
  &=  C_N(a_1,a_2,...,a_N) \int_{\Gamma} dz_j \frac{1}{z_j}
  [P(0,...,0,z_j,0,...,0)]^{\frac{D-L-E-1}{2}} \nonumber\\
  &+ C_{N-1,\hat{j}}(a_1,a_2,...,a_N)\int_{\Gamma} dz_j 
  [P(0,...,0,z_j,0,...,0)]^{\frac{D-L-E-1}{2}}.
  \label{eq:no GF expand}
\end{align}
The new unknown reduction coefficient $C_{N-1, \hat{j}}$ corresponds to a master integral where only $z_j$ has a power of zero, while all others have a power of one. The reduction coefficient $C_{N}$ is known from the previous formula. The remaining integral is a definite integral in $z_j$, with the integration region constrained to the physical domain, which is typically the region where $P(0, \dots, 0, z_j, 0, \dots, 0) \geq 0$.  Similarly, after computing the reduction coefficients for all sub top sector, we can take the residues of the $N-2$ Baikov variables, excluding $z_i$ and $z_j$, to obtain:
\begin{align}
    I_0 = C_NI_N + C_{N-1,\hat{i}}I_{N-1,\hat{i}} + C_{N-1,\hat{j}}I_{N-1,\hat{j}} + C_{N_2,\hat{i}\hat{j}}I_{N-2,\hat{i}\hat{j}},
    \label{eq:GF n-2 01}
\end{align}
where $I_{0/N/N-1/N-2}$ represents a series of double integrals, with the integration region $\Gamma$ defined by $P(0, \dots, 0, z_i, 0, \dots, 0, z_j, 0, \dots, 0) \geq 0$. For example, the expression for $I_{N-1,\hat{i}}$ is:
\begin{small}
    \begin{align}
I_0 &= \int_{\Gamma} dz_idz_j
\frac{1}{z_i^{a_i}z_j^{a_j}} \frac{1 }{ (a_{1}-1)! \dots (a_{i-1}-1)! (a_{i+1}-1)! \dots (a_{j-1}-1)! (a_{j+1}-1)! \dots (a_N-1)! } \nonumber\\
\times & \Big(\frac{\partial}{\partial z_{1}}\Big)^{a_{1}-1} \dots
\Big(\frac{\partial}{\partial z_{i-1}}\Big)^{a_{i-1}-1}
\Big(\frac{\partial}{\partial z_{i+1}}\Big)^{a_{i+1}-1} \dots
\Big(\frac{\partial}{\partial z_{j-1}}\Big)^{a_{j-1}-1} 
\Big(\frac{\partial}{\partial z_{j+1}}\Big)^{a_{j+1}-1} \dots
\Big(\frac{\partial}{\partial z_{N}}\Big)^{a_{N}-1} \nonumber\\
\times &  [P(\vec{z})]^{\frac{D-L-E-1}{2}}
\Big|_{z_1,\dots,z_{i-1},z_{i+1},\dots,z_{j-1},z_{j+1},\dots,z_N=0},\\
I_{N} &= \int_{\Gamma} dz_i dz_j \frac{1}{z_iz_j}
[P(0,...,0,z_i,0,...,0,z_j,0,...,0)]^{\frac{D-L-E-1}{2}} ,\\
I_{N-1,\hat{i}} &= \int_{\Gamma} dz_i dz_j \frac{1}{z_j}
[P(0,...,0,z_i,0,...,0,z_j,0,...,0)]^{\frac{D-L-E-1}{2}}, \\
I_{N-1,\hat{j}} &= \int_{\Gamma} dz_i dz_j \frac{1}{z_i}
[P(0,...,0,z_i,0,...,0,z_j,0,...,0)]^{\frac{D-L-E-1}{2}} ,\\
I_{N-2,\hat{i}\hat{j}} &= \int_{\Gamma} dz_i dz_j 
[P(0,...,0,z_i,0,...,0,z_j,0,...,0)]^{\frac{D-L-E-1}{2}}. 
\label{eq:GF n-2 02}
\end{align}
\end{small}
Hence, only reduction coefficient \( C_{N-2,\hat{i}\hat{j}} \) is unknown, with all other reduction coefficients $C_{N/N-1}$ already determined. As long as we have the results for all  double integrals above, we can determine the reduction coefficients.

Using the approach described above, we iteratively reduce the number of Baikov variables for which we take residues, until only one Baikov variable remains. At this point, the corresponding integral becomes an $(N-1)$-fold integral. Once the results for all integrals are available, we can determine all the reduction coefficients.

\subsection{Examples} \label{sec: 2.3}
Here we demonstrate a simple example to illustrate. Considering a massive bubble integral:
\begin{equation}
    I(a_1,a_2)=\int_{\Gamma} \frac{d^Dl}{(l^2-m^2)^{a_1}[(l+q)^2-m^2]^{a_2}},
\end{equation}
the corresponding Baikov polynomial is: 
\begin{equation}
    P(z_1,z_2) = -\frac{1}{4}(z_1-z_2)^2+\frac{1}{2}q^2(z_1+z_2)-\frac{1}{4}q^4+q^2m^2.
\end{equation}
The master integrals are $I(1,1),I(1,0),I(0,1)$, then the expansion formula is: 
\begin{align}
     \int_{\Gamma}\frac{dz_1dz_2}{z_1^{2}z_2}P(z_1,z_2)^{\frac{D-3}{2}}
    = & C_{2}(2,1) 
    \int_{\Gamma}\frac{dz_1dz_2}{z_1z_2}P(z_1,z_2)^{\frac{D-3}{2}}
    + C_{1,\hat{1}}(2,1)
    \int_{\Gamma}\frac{dz_1dz_2}{z_2}P(z_1,z_2)^{\frac{D-3}{2}}\nonumber\\
    + & C_{1,\hat{2}}(2,1)
    \int_{\Gamma}\frac{dz_1dz_2}{z_1}P(z_1,z_2)^{\frac{D-3}{2}}.
\end{align}
Taking the residues of $z_1,z_2$ at the origin, we have 
\begin{equation}
  C_2(2,1) = [P(0,0)]^{-\frac{D-3}{2}}
  \frac{\partial}{\partial z_1}[P(z_1,z_2)]^{\frac{D-3}{2}}\Big|_{z_1,z_2=0}
  = \frac{D-3}{4m^2-q^2}.
\end{equation}
Taking only the residue of $z_2$ at the origin, we have:
\begin{small}
    \begin{align}
  C_{1,\hat{2}}(2,1) =
  \frac{
  \int_{z_1^-}^{z_1^+}\frac{dz_1}{z_1^{2}}P(z_1,0)^{\frac{D-3}{2}}
  -C_2(2,1)
  \int_{z_1^-}^{z_1^+}\frac{dz_1}{z_1}P(z_1,0)^{\frac{D-3}{2}}
  }{
  \int_{z_1^-}^{z_1^+}dz_1P(z_1,0)^{\frac{D-3}{2}}
  }
  = -\frac{D-2}{2m^2(4m^2-q^2)}.
\end{align}
\end{small}
The integral with respect to $z_1$ is a definite integral, with the integration region ensuring $P(z_1,0)>0$. Since $P(z_1,0)$ is a quadratic function of $z_1$, the upper and lower limits of integration $z_1^+,z_1^-$ are. 
\begin{equation}
    z_1^{\pm} = q^2 \pm 2\sqrt{m^2q^2}.
\end{equation}

Next, we provide an example of a sunset diagram
\begin{equation}
    I_{Sunset}(a_1,a_2,a_3) = \int \frac{d^Dl_1d^Dl_2}{(l_1^2-m_1^2)^{a_1}(l_2^2-m_2^2)^{a_2}[(l_1+l_2+q)^2-m_3^2]^{a_3}}.
    \label{eq:sunset 01}
\end{equation}
In this example, $L=2$ and $E=1$, resulting in a total of five independent Lorentz scalar products involving loop momenta: $l^2_1$, $l^2_2$, $(l_1\cdot l_2)$, $(l_1\cdot q)$, and $(l_2\cdot q)$. However, there are only three propagators, so we need to extend them to five:
\begin{align}
    I(a_1,a_2,a_3,a_4,a_5) &= \int \frac{d^Dl_1d^Dl_2}{(l_1^2-m_1^2)^{a_1}(l_2^2-m_2^2)^{a_2}[(l_1+l_2+q)^2-m_3^2]^{a_3}[(l_1+q)]^2]^{a_4}[(l_2-q)^2]^{a_5}} \nonumber\\
    &= \int_{\Gamma} \frac{dz_1dz_2dz_3dz_4dz_5}{z_1^{a_1}z_2^{a_2}z_3^{a_3}z_4^{a_4}z_5^{a_5}}P(\vec{z})^{\frac{D-4}{2}}.
\end{align}
Then sunset integral \eqref{eq:sunset 01} can be expressed as
\begin{equation}
    I_{Sunset}(a_1,a_2,a_3) = I(a_1,a_2,a_3,0,0).
\end{equation}
There are seven master integrals in this family:
\begin{align}
    I^1 &= I(2,1,1,0,0),\quad I^2 = I(1,1,1,-1,0),\quad I^3 = I(1,1,1,0,-1), \quad I^4 = I(1,1,1,0,0),\nonumber \\
    I^5 &= I(0,1,1,0,0),\quad I^6 = I(1,0,1,0,0), \quad I^7 = I(1,1,0,0,0).\nonumber
\end{align}
The expansion of the sunset integral is
\begin{equation}
    I(a_1,a_2,a_3,0,0) = \sum_{n=1}^7 C_n(a_1,a_2,a_3,0,0) I^n.
\end{equation}
By taking residues at the seven combinations: $z_1=z_2=z_3=0$, $z_1=z_2=0$, $z_1=z_3=0$, $z_2=z_3=0$, $z_1=0$, $z_2=0$, and $z_3=0$, we obtain a set of seven linear equations, all of which are independent.
 \begin{align}
     I_{ijk}^0=\sum_{n=1}^7 C_n(a_1,a_2,a_3,0,0) I^n_{ijk}  ,\quad  i,j,k=0/1.
     \label{eq:sunset eq}
 \end{align}
The indices $i$, $j$, and $k$ are used to label the 7 groups  of definite  integrals, where each can take a value of 0 or 1, with the condition that not all are 0 at the same time. These indices correspond to the Baikov variables $z_1$, $z_2$, and $z_3$. A value of 1 means the residue at the origin of the corresponding $z$ is taken, while a value of 0 means it is not. For example, $ijk = 110$,  the corresponding integral is
\begin{small}
     \begin{align}
     I_{110}^0 &= \int_{\Gamma} \frac{dz_3dz_4dz_5}{z_3^{a_3}z_4^{a_4}z_5^{a_5}}
     \frac{1}{(a_1-1)!(a_2-1)!}
     \Big(\frac{\d}{\d z_1}\Big)^{a_1-1} \Big(\frac{\d}{\d z_2}\Big)^{a_2-1}
     [P(\vec{z})]^{\frac{D-4}{2}}\Big|_{z_1=z_2=0}, \\
     I_{110}^1 &=  \int_{\Gamma} dz_3dz_4dz_5\frac{1}{z_3}\frac{\d}{\d z_1}
     [P(\vec{z})]^{\frac{D-4}{2}}\Big|_{z_1=z_2=0}, \quad
     I_{110}^2 =  \int_{\Gamma} dz_3dz_4dz_5\frac{z_4}{z_3}
     [P(\vec{z})]^{\frac{D-4}{2}}\Big|_{z_1=z_2=0}, \\
     I_{110}^3 &=  \int_{\Gamma} dz_3dz_4dz_5\frac{z_5}{z_3}
     [P(\vec{z})]^{\frac{D-4}{2}}\Big|_{z_1=z_2=0}, \quad
     I_{110}^4 =  \int_{\Gamma} dz_3dz_4dz_5\frac{1}{z_3}
     [P(\vec{z})]^{\frac{D-4}{2}}\Big|_{z_1=z_2=0}, \\
     I_{110}^6 &=  \int_{\Gamma} dz_3dz_4dz_5
     [P(\vec{z})]^{\frac{D-4}{2}}\Big|_{z_1=z_2=0}, \quad
      I_{110}^5 = I_{110}^7 =0.
 \end{align}
\end{small}
The system in \eqref{eq:sunset eq}, comprising seven linearly independent equations for seven unknown reduction coefficients, is in principle straightforward to solve. The primary difficulty of this method, however, lies not in solving the system but in its construction. Specifically, the coefficients of the linear system are derived from a set of integrals whose explicit evaluation is often the most challenging part. While theoretically sound, the practical implementation of this approach is nontrivial due to the computational difficulty of these integrals.

\section{Generating Function in Baikov Representation}\label{sec: 3}

With the introduction to the Baikov representation above, it is straightforward to extend it to the generating function framework. First, we construct a generating function of the form \eqref{eq:GF01} for $L$-loop Feynman integral \eqref{eq:tradition form}:
\begin{small}
    \begin{equation}
\int \prod_{j=1}^L d^D l_j \frac{1}{(D_1-t_1) (D_2-t_2) \cdots (D_N-t_N)}=\sum_{a_1,\cdots,a_N=1}^{\infty}\int \prod_{j=1}^L d^D l_j \frac{t_1^{a_1-1}\cdots t_N^{a_N-1}}{D_1^{a_1} D_2^{a_2} \cdots D_N^{a_N}}.
\label{eq:GF 03}
\end{equation}
\end{small}
The corresponding expansion into master integrals is as follows:
\begin{equation}
\begin{aligned}
    \int \prod_{j=1}^L d^D l_j \frac{1}{(D_1-t_1)  \cdots (D_N-t_N)}=\sum_i GF_i(t_1,\cdots,t_N)\int  \prod_{j=1}^L d^D l_j \frac{1}{D_1^{a^i_1} \cdots D_N^{a^i_N}}.
\end{aligned}
\end{equation}
Compared to the general form of Feynman integral expansions into master integrals, here the reduction coefficients are replaced by generating functions:
\begin{equation}
GF_i(t_1,t_2,\cdots,t_N)=\sum_{a_1,\cdots,a_N=1}^{\infty}C_i(a_1,a_2,\cdots,a_N)t_1^{a_1-1}t_2^{a_2-1}\cdots t_N^{a_N-1}.
\end{equation}
In fact, the introduction of the parameter $t_i$ in our generating function is essentially equivalent to performing a mass shift on the corresponding propagator. The idea that derivatives with respect to the mass can be used to construct reduction coefficients with higher-order poles has already appeared in earlier works (see, e.g, \cite{Tarasov:1998nx}) and is not new. Our work is to cast this idea into a systematic generating function framework, yielding more compact expressions.
 As mentioned in the introduction, we expect that for studying the structure of reduction coefficients involving higher-order poles, the generating function may provide a simpler and more unified description.
If we want to use the method from Section \ref{sec: 2} to compute the generating function, we need to address two key points. First, we need to determine the Baikov representation of the left-hand side of \eqref{eq:GF 03}. Second, we must identify how the operation of taking residues at $z_i = 0$, as described in Section \ref{sec: 2}, should be adapted in this context. 

For the first point, there are two possible approaches: the first is to keep the auxiliary parameter $t_i$ within the propagators, and the second is to treat the entire propagator $D_i - t_i$ as a new Baikov variable $z_i$, with the auxiliary parameter $t_i$ appearing correspondingly in the Baikov polynomial $P(\vec{z})$ (i.e., the measure). Expressed mathematically, \red{these} can be written as:
\begin{align}
    \operatorname{Type-1}:& \quad \int_{\Gamma} \frac{d^Nz}{(z_1-t_1)(z_2-t_2)\dots(z_N-t_N)}[P(z_1,z_2,\dots,z_N)]^{\frac{D-L-E-1}{2}},  
    \label{eq:GF type1}
    \\
    \operatorname{Type-2}:& \quad \int_{\Gamma} \frac{d^Nz}{z_1 z_2 \dots z_N}[P(z_1+t_1,z_2+t_2,\dots,z_N+t_N)]^{\frac{D-L-E-1}{2}}.  
    \label{eq:GF type2}
\end{align}
With these two types of Baikov representations, let us now examine how the second step, 'taking residues', should be adapted. First, for the Type-I representation, we cannot naively take the residue at $z_i = 0$. To illustrate this point, let us consider a Tadpole diagram as an example. In this case there is only one master integral, the expansion is 
\begin{equation}
    \int dz \frac{[P(z)]^{\frac{D-2}{2}}}{z-t}=GF(t) \int dz \frac{[P(z)]^{\frac{D-2}{2}}}{z},
\end{equation}
this function is non-singular at $z = 0$. If we naively take the residue at $z = 0$ on both sides, the left-hand side vanishes because it is non-singular at $z = 0$, while the right-hand side reduces to $GF(t)\cdot [P(0)]^{\frac{D-2}{2}}$. Clearly, the two sides are not equal. This discrepancy arises because, when constructing the generating function, it is actually represented as an infinite series expansion:
\begin{equation}
    \int dz \frac{[P(z)]^{\frac{D-2}{2}}}{z-t}\equiv \sum_{n=1}^\infty \int dz \frac{[P(z)]^{\frac{D-2}{2}}\cdot t^{n-1}}{z^n}.
    \label{eq:GF 101}
\end{equation}
Therefore, the operation of taking the residue at $z = 0$ should be performed separately for each term in the infinite series, followed by summing the results.
\begin{footnotesize}
    \begin{equation}
         Res_{z\to 0}\Bigg\{\sum_{n=1}^\infty \int dz \frac{[P(z)]^{\frac{D-2}{2}}\cdot t^{n-1}}{z^n}\Bigg\}=\sum_{n=1}^\infty \Bigg\{\frac{t^{n-1}}{(n-1)!}\frac{\partial^{n-1}}{\partial z^{n-1}}\left([P(z)]^{\frac{D-2}{2}}\right)\Bigg|_{z=0}\Bigg\}=[P(t)]^{\frac{D-2}{2}}.
         \label{eq:GF series}
    \end{equation}  
\end{footnotesize}
Since $[P(t)]^{\frac{D-2}{2}}$ is simply a power of a polynomial, the final equality in the above expression always converges as a Taylor series when $t$ is sufficiently small (but not zero). However, the equality in \eqref{eq:GF 101} holds only under the condition $|z| > |t| > 0$. Therefore, taking the residue at $z = 0$ on both sides of \eqref{eq:GF 101} to establish equations is not valid. It is easy to see that the result of \eqref{eq:GF series} corresponds to taking the residue of the left-hand side of \eqref{eq:GF 101} at $z = t$. This is intuitive, as \eqref{eq:GF 101} holds in the region $|z| > t$. Therefore, if we perform a contour integral along $|z| = t + \epsilon$ on both sides of the equation, the left-hand side diverges at $z = t$, while the right-hand side diverges at $z = 0$. Ultimately, this also leads to the result in \eqref{eq:GF series}. On the other hand, if we directly take the residue at $z_i = 0$ for the second form of the Baikov representation \eqref{eq:GF type2}, we can obtain the same result. In summary, when computing the generating function for these two different forms of the Baikov representation, different methods should be used. For the first form, a contour integral along $|z_i| = t_i + \epsilon$ should be performed, while for the second form, the residue at $z_i = 0$ can be taken directly. For the remaining definite integral part, both methods yield the same result. For example for a bubble diagram, with the two forms of the Baikov representation given as:

\begin{equation}
    \begin{aligned}
        &\int \frac{P(z_1,z_2)^{\frac{D-3}{2}}dz_1dz_2}{(z_1-t_1)(z_2-t_2)}
        \stackrel{z_1\rightarrow t_1}{\longrightarrow} 
        \int_{\alpha}^{\beta}\frac
        {P(t_1,z_2)^{\frac{D-3}{2}}dz_2}{z_2-t_2},\\
        &\int \frac{P(z_1+t_1,z_2+t_2)^{\frac{D-3}{2}}dz_1dz_2}{z_1z_2}
        \stackrel{z_1\rightarrow 0}{\longrightarrow}
        \int_{\alpha'}^{\beta'}\frac
        {P(t_1,z_2+t_2)^{\frac{D-3}{2}}dz_2}{z_2},
    \end{aligned}
\end{equation}
where, the integration limits $\a$ and $\b$ are the two roots of $P(t_1,z_2) = 0$, while $\a'$ and $\b'$ are the two roots of $P(t_1,z_2+t_2) = 0$. Hence, we have $\a'+t_2=\a$, $\b'+t_2=\b$. It can be observed that the two integrals differ only by a shift in $z_2$, so they are equivalent. In subsequent derivations, we adopt the second form of expression, treating $D_i - t_i$ as a new Baikov variable $z_i$ on the left-hand side of the equation.

\section{Methodological Framework}\label{sec: 4}

We start our discussion from the following formula:
\begin{equation}
  \int_{\Gamma} \frac{d^Nz}{z_1 z_2 \dots z_N}
  [P( \vec{z} + \vec{t} )]^{\gamma}
  = \sum_i GF_i(\vec{t}) \int \frac{d^Nz}{z_1^{a^i_1} \cdots z_N^{a^i_N}}
  [P( \vec{z} )]^{\gamma},
  \label{eq: generation function of Baikov}
\end{equation}
where, we use $\gamma=\frac{D-L-E-1}{2}$ to simplify the expression.  If we only focus on the case where $a_i > 0$ and $N = \frac{L(L+1)}{2} + LE$, we can then follow the method outlined in Section \ref{sec: 2}  to establish the relations satisfied by the generating functions and compute them iteratively. More specifically, we first take the residues of all $N$ Baikov variables at $z_i = 0$. On the right-hand side of the equation, only the master integral with $a_1 = a_2 = \cdots = a_N = 1$ remains, as all other master integrals vanish due to the absence of poles. As a result, we can immediately obtain:
\begin{equation}
    [P(\vec{t})]^{\gamma} = GF_N(\vec{t})\cdot [P(\vec{0})]^{\gamma}\Rightarrow GF_N(\vec{t})=\left(\frac{P(\vec{t})}{P(\vec{0})}\right)^{\gamma}.
    \label{eq:GFn}
\end{equation}
Then the reduction coefficients $ C_N(a_1,a_2,\cdots,a_N)$ are:
\begin{equation}
    C_N (a_1,a_2,\dots ,a_N) = \frac{[P(\vec{0})]^{-\gamma}}{(a_1-1)!\cdots (a_N-1)!}
    \left(\frac{\partial}{\partial t_1}\right)^{a_1-1}\dots
  \left(\frac{\partial}{\partial t_{N}}\right)^{a_{N}-1}
  [P(\vec{t})]^{\gamma}
  \Bigg|_{\vec{t}=\vec{0}}.
  \label{eq:nton}
\end{equation}
The result for the top sector, as expressed in \eqref{eq:GFn} \eqref{eq:nton}, is well known and can be derived straightforwardly within the Baikov representation. It follows from the fact that the leading coefficient corresponds to the maximal-cut residue of the Baikov polynomial. In the second step, we take residues at $z_i = 0$ for $N-1$ of the $z_i$'s, excluding $z_j$, and obtain:
\begin{align}
    \int_{\tilde{\Gamma}} \mathrm{d}z_j &\frac{[P(t_1,\dots ,t_{j-1}, z_j+t_j, t_{j+1},\dots,t_N)]^{\gamma}}{z_j} =
    GF_N(\vec{t}) \int_{\Gamma} \mathrm{d}z_j \frac{[P(0,\dots ,z_j,\dots,0)]^{\gamma}}{z_j} \nonumber \\
    &+ GF_{N-1,\hat{j}}(\vec{t})  \int_{\Gamma} \mathrm{d}z_j [P(0,\dots ,z_j,\dots,0)]^{\gamma},
    \label{eq:GF n to n-1 01}
\end{align}
where $GF_N(\vec{t})$ is already given by \eqref{eq:GFn}. The integration regions $\Gamma$ and $\tilde{\Gamma}$ are defined as the regions where the inequalities  
$P(0, \dots, z_j, \dots, 0) \geq 0$ and  
$P(t_1, \dots, t_{j-1}, z_j + t_j, t_{j+1}, \dots, t_N) \geq 0$, respectively, are satisfied.  
In most cases, the Baikov polynomial is quadratic in $z_j$, with a negative coefficient for the quadratic term, which ensures that $P(z_j)$ is concave. To be detailed, the Baikov polynomial can be expressed as:
\begin{align}
     &P(0,\dots ,z_j,\dots,0) = -C(z_j-z_j^+)(z_j-z_j^-), \\
     &P(t_1,\dots ,t_{j-1}, z_j+t_j, t_{j+1},\dots,t_N) =  -C’(\vec{t})(z_j-\tilde{z}_j^+(\vec{t}))(z_j-\tilde{z}_j^-(\vec{t})),
\end{align}
where $C$ and $C'(\vec{t})$ are positive numbers (with $C'(\vec{t}) > 0$ even when it depends on $t_i$, since these $t_i$ can be taken sufficiently small so as not to affect the sign of $C'(\vec{t})$). Here, $z_j^{\pm}$ denote the two roots of the equation $P(0, \dots, z_j, \dots, 0) = 0$, while $\tilde{z}_j^{\pm}(\vec{t})$ are the two roots of $P(t_1, \dots, t_{j-1}, z_j + t_j, t_{j+1}, \dots, t_N) = 0$. These roots define the boundaries of the integration regions. Next, we need to handle the integrals in \eqref{eq:GF n to n-1 01}. The key to computing these integrals lies in identifying their shared structure. we have
\begin{align}
    I_1(z_j^{-},z_j^+) & = \int_{z_j^{-}}^{z_j^+} dz_j 
    \frac{\big( C(z_j-z_j^{-})(z_j-z_j^+) \big)^{\gamma}}{z_j} \nonumber\\
    &= C^{\gamma} \frac{(z_j^+-z_j^{-})^{2\gamma+1}}{z_j^{-}}
    \frac{\Gamma(\gamma+1)^2}{\Gamma(2\gamma+2)}
    \cdot \sideset{_2}{_1}{\mathop{\mathrm{F}}}\left( \begin{array}{c}  1,\gamma+1 \\ 2\gamma+2 \end{array} \middle| 1-\frac{z_j^{+}}{z_j^-} \right), \label{eq:integral 1}\\
    I_2(z_j^{-},z_j^+) & = \int_{z_j^{-}}^{z_j^+} dz_j 
    \big( C(z_j-z_j^{-})(z_j-z_j^+) \big)^{\gamma} =
    C^{\gamma} \Big( \frac{z_j^+-z_j^{-}}{2} \Big)^{2\gamma+1}
    \frac{\Gamma(\frac{1}{2})\Gamma(\gamma+1)}{\Gamma(\gamma+\frac{3}{2})},
    \label{eq:integral 2}
\end{align}
where $\sideset{_2}{_1}{\mathop{\mathrm{F}}}\left( \begin{array}{c} a,b \\ c \end{array} \middle| z \right)$ is the hypergeometric function, defined as 
\begin{equation}
    \sideset{_2}{_1}{\mathop{\mathrm{F}}}\left( \begin{array}{c} a,b \\ c \end{array} \middle| z \right) = \sum_{n=0}^{\infty} \frac{(a)_n(b)_n}{n!(c)_n}z^n.
\end{equation}
Here, $(x)_n$ is Pochhammer’s Symbol defined as $(x)_n=\Gamma(x+n)/\Gamma(x)$. $\Gamma(x)$ is gamma function. As long as $P(\vec{z})$ is a known quadratic form, the values of $z_j^{\pm}$ and $\tilde{z}_j^{\pm}$ can be easily determined. Using these values, we can express $GF_{N-1;\hat{j}}(\vec{t})$ as:  
\begin{equation}  
    GF_{N-1,\hat{j}} (\vec{t}) = \frac{I_1(\tilde{z}_j^-(\vec{t}),\tilde{z}_j^+(\vec{t})) - GF_N(\vec{t}) I_2(z_j^-,z_j^+)}{I_1(z_j^-,z_j^+)}. 
    \label{eq:GF n to n-1}
\end{equation}  
The generating functions for reductions to lower topologies can also be derived iteratively in a similar manner. 

Since the generating function serves merely as an auxiliary tool. During the differentiation process to compute the reduction coefficients, we need apply the linear transformation of the hypergeometric function: 
\begin{equation}
\begin{aligned}
& \sideset{_2}{_1}{\mathop{\mathrm{F}}}\left(\begin{array}{c}
1 , \gamma +1 \\
2\gamma +2
\end{array} \Bigg | 1-\frac{z_j^{+}}{z_j^{-}}\right)=
\Bigg( \frac{z_j^{+}}{z_j^{-}} \Bigg)^{-\gamma-1}
\sideset{_2}{_1}{\mathop{\mathrm{F}}}\left(\begin{array}{c}
2\gamma +1 ,\gamma + 1\\
2\gamma +2
\end{array} \Bigg | 1-\frac{z_j^{-}}{z_j^{+}}\right).
\end{aligned}
\label{eq:trans 1}
\end{equation}
This transformation ensures that the parameters of the hypergeometric function remain unchanged during differentiation: 
\begin{small}
      \begin{equation}
    \frac{d}{dz}  \sideset{_2}{_1}{\mathop{\mathrm{F}}}\left(\begin{array}{c}
    2\gamma +1 ,\gamma + 1\\  2\gamma +2
    \end{array} \Bigg | z\right)
    = \frac{2\gamma+1}{z}\left\{(1-z)^{-1-\gamma}- \sideset{_2}{_1}{\mathop{\mathrm{F}}}\left(\begin{array}{c}
    2\gamma +1 ,\gamma + 1\\  2\gamma +2
    \end{array} \Bigg | z\right)\right\} .
    \label{eq:trans 2}
\end{equation}
\end{small}
Consequently, the final reduction coefficients only involve this single type of hypergeometric function. Moreover, it can be verified that the coefficients in front of the hypergeometric function cancel out, leaving only the reduction coefficients. Another special case arises when $z_j^+ = 0$, where the above transformation fails. In this case, the following substitution can be used instead.
\begin{equation}
  \sideset{_2}{_1}{\mathop{\mathrm{F}}}\left( \begin{array}{c} a,b\\c \end{array} \middle| 1 \right)
  = \frac{\Gamma(c)\Gamma(c-a-b)}{\Gamma(c-a)\Gamma(c-b)}.
  \label{eq:trans 3}
\end{equation}
Specific examples will be provided in Section \ref{sec: 7}.

For the case where $N < \frac{L(L+1)}{2} + LE$, we can similarly follow the operations in Section \ref{sec: 2.3}. By introducing additional propagators, all scalar products involving loop momenta can be expressed as linear combinations of the propagators. The powers of these additional propagators are then set to zero. Relevant examples will be provided in Section \ref{sec: 5}.



\section{One loop examples}\label{sec: 5}
In this section, we present several simple one-loop examples to illustrate how the proposed method can be applied.

\subsection{Tadpole}\label{sec: 5.1}
We begin our discussion with the simplest example of the tadpole integral, which has only one propagator. 
\begin{equation}
  I(a) = \int \frac{d^Dl}{(l^2-m^2)^a} 
  \rightarrow \int_{\Gamma} \frac{dz}{z^a} 
  P(z)^{\frac{D-2}{2}} ,\quad
  P(z) = z+m^2.
\end{equation}
This integral has only one master integral, $I(1)$. The expansion of the generating function over the master integral can be written as:
\begin{equation}
  \int_{\Gamma} \frac{dz}{z} P(z+t)^{\frac{D-2}{2}} =
  GF(t) \int_{\Gamma} \frac{dz}{z} P(z)^{\frac{D-2}{2}}.
\end{equation}
Taking the residue on both sides of the above equation at $z = 0$, we have
\begin{equation}
  P(t)^{\frac{D-2}{2}} = GF(t) P(0)^{\frac{D-2}{2}}.
\end{equation}
Then the reduction coefficients are:
\begin{align}
    C(a) =& \frac{1}{(a-1)!} \Big( \frac{\partial}{\partial t} \Big)^{a-1} GF(t) 
  = \frac{(m^2)^{(2-D)/2}}{(a-1)!} \Big( \frac{\partial}{\partial t} \Big)^{a-1}
  [t+m^2]^{\frac{D-2}{2}}\Big|_{t=0} \nonumber\\
  =& \frac{(-1)^{a-1}(1-\frac{D}{2})_{a-1}}{(a-1)!(m^2)^{a-1}}.
\end{align}

\subsection{Massive bubble}\label{sec: 5.2}
Next, we present a less trivial yet typical example: the massive bubble integral.
\begin{equation}
  I(a_1,a_2) = \int \frac{d^Dl}{(l^2-m_1^2)^{a_1} [(l+p)^2-m_2^2]^{a_2}}
\rightarrow \int_\Gamma \frac{dz_1 dz_2}{z_1^{a_1} z_2^{a_2}} P(z_1,z_2)^{\frac{D-3}{2}} ,
\end{equation}
where $L=1,E=1$ and $z_1=l^2-m_1^2$, $z_2=(l+p)^2-m_2^2$. The Baikov Polynomial is
\begin{equation}
    P(z_1,z_2) = \begin{vmatrix}
l^2 & l\cdot p \\
l\cdot p & p^2
\end{vmatrix} 
= -\frac{1}{4} (z_2-z_1+m_2^2-m_1^2-p^2)^2 + p^2(z_1+m_1^2).
\end{equation}
This integral has three master integrals $I(1,1)$, $I(1,0)$, $I(0,1)$. The expansion of generating function is expressed as
\begin{align}
  &\int_{\Gamma'} \frac{dz_1 dz_2}{z_1z_2} P(z_1+t_1, z_2+t_2)^{\frac{D-3}{2}}
  = GF_2(\vec{t}) \int_\Gamma \frac{dz_1 dz_2}{z_1 z_2} P(z_1,z_2)^{\frac{D-3}{2}} \nonumber\\
  &+ GF_{1,\hat{1}}(\vec{t}) \int_\Gamma \frac{dz_1 dz_2}{z_2} P(z_1,z_2)^{\frac{D-3}{2}}
  + GF_{1,\hat{2}}(\vec{t}) \int_\Gamma \frac{dz_1 dz_2}{z_1} P(z_1,z_2)^{\frac{D-3}{2}}.
\end{align}
Taking the residue of $z_1,z_2$ at $z_i = 0$, we have 
\begin{equation}
  P(t_1,t_2)^{\frac{D-3}{2}} = GF_2(\vec{t}) P(0,0)^{\frac{D-3}{2}}.
\end{equation}
Then, we take the residue at $z_1 = 0$ , obtaining
\begin{small}
    \begin{equation}
  \int_{\tilde{z}_2^-(\vec{t})}^{\tilde{z}_2^+(\vec{t})} \frac{dz_2}{z_2} P(t_1, z_2+t_2)^{\frac{D-3}{2}}
  = GF_2(\vec{t}) \int_{z_2^-}^{z_2^+} \frac{dz_2}{z_2} P(0,z_2)^{\frac{D-3}{2}}
  + GF_{1,\hat{2}}(\vec{t}) \int_{z_2^-}^{z_2^+} dz_2 P(0,z_2)^{\frac{D-3}{2}},
\end{equation}
\end{small}
where $\tilde{z}_2^{\pm}(\vec{t})$ are the two roots of $P(t_1,z_2+t_2)=0$ , $z_2^{\pm}$ are the two roots of $P(0,z_2)= 0$, 
\begin{align}
  \tilde{z}_2^{\pm}(\vec{t}) &= t_1-t_2+m_1^2-m_2^2+p^2 \pm 2\sqrt{p^2(t_1+m_1^2)}, \\
  z_2^{\pm} &= m_1^2-m_2^2+p^2 \pm 2\sqrt{p^2m_1^2}.
\end{align}
$GF_2(\vec{t})$ has been calculated above. By \eqref{eq:integral 1}, \eqref{eq:integral 2}, \eqref{eq:GF n to n-1}, the expression for $GF_{1,\hat{2}}(\vec{t})$ is given by 
\begin{align}
  GF_{1,\hat{2}}(\vec{t}) =& 
  \frac{\Gamma(\frac{D-1}{2}) \Gamma(\frac{D}{2})}{\Gamma(\frac{1}{2})\Gamma(D-1)}
  \Bigg[
  \Bigg(\frac{2(\tilde{z}_2^+ (\vec{t})- \tilde{z}_2^-(\vec{t}))}{z_2^+-z_2^-} \Bigg)^{D-2}
  \frac{1}{\tilde{z}_2^-(\vec{t})}
  \sideset{_2}{_1}{\mathop{\mathrm{F}}}\left( \begin{array}{c} 1,\frac{D-1}{2} \\ D-1 \end{array} \middle| 1-\frac{\tilde{z}_2^+(\vec{t})}{\tilde{z}_2^-(\vec{t})}  \right) \nonumber\\
  &\qquad - \Bigg(\frac{P(t_1,t_2)}{P(0,0)}\Bigg)^{\frac{D-3}{2}}
  \frac{2^{D-2}}{z_2^-}
  \sideset{_2}{_1}{\mathop{\mathrm{F}}}\left( \begin{array}{c} 1,\frac{D-1}{2} \\ D-1 \end{array} \middle| 1-\frac{z_2^+}{z_2^-}  \right)
  \Bigg]\nonumber\\
  =&
  \Bigg(\frac{(\tilde{z}_2^+ (\vec{t})- \tilde{z}_2^-(\vec{t}))}{z_2^+-z_2^-} \Bigg)^{D-2}
  \frac{1}{z_2^-}
  \sideset{_2}{_1}{\mathop{\mathrm{F}}}\left( \begin{array}{c} 1,\frac{D-1}{2} \\ D-1 \end{array} \middle| 1-\frac{\tilde{z}_2^+(\vec{t})}{\tilde{z}_2^-(\vec{t})}  \right) \nonumber\\
  &\qquad - \Bigg(\frac{P(t_1,t_2)}{P(0,0)}\Bigg)^{\frac{D-3}{2}}
  \frac{1}{z_2^-}
  \sideset{_2}{_1}{\mathop{\mathrm{F}}}\left( \begin{array}{c} 1,\frac{D-1}{2} \\ D-1 \end{array} \middle| 1-\frac{z_2^+}{z_2^-}  \right).
\end{align}
The second equality in the above expression makes use of the Legendre duplication formula:

\begin{equation}
    \Gamma(2z)\Gamma(\frac{1}{2})=2^{2z-1}\Gamma(z)\Gamma(z+\frac{1}{2}).
\end{equation}
Similarly, we can obtain the generating function for the reduction to another tadpole master integral.
\begin{align}
  GF_{1,\hat{1}}(\vec{t}) &= 
  \Bigg(\frac{(\tilde{z}_1^+ (\vec{t})- \tilde{z}_1^-(\vec{t}))}{z_1^+-z_1^-} \Bigg)^{D-2}
  \frac{1}{\tilde{z}_1^-(\vec{t})}
  \sideset{_2}{_1}{\mathop{\mathrm{F}}}\left( \begin{array}{c} 1,\frac{D-1}{2} \\ D-1 \end{array} \middle| 1-\frac{\tilde{z}_1^+(\vec{t})}{\tilde{z}_1^-(\vec{t})}  \right) \nonumber\\
  &\qquad - \Bigg(\frac{P(t_1,t_2)}{P(0,0)}\Bigg)^{\frac{D-3}{2}}
  \frac{1}{z_1^-}
  \sideset{_2}{_1}{\mathop{\mathrm{F}}}\left( \begin{array}{c} 1,\frac{D-1}{2} \\ D-1 \end{array} \middle| 1-\frac{z_1^+}{z_1^-}  \right),
\end{align}
where $\tilde{z}_1^{\pm}(\vec{t})$ are the two roots of $P(z_1+t_1,t_2)=0$ , $z_1^{\pm}$ are the two roots of $P(z_1,0)= 0$, 
\begin{align}
    \tilde{z}_1^{\pm} (\vec{t})&= t_2 - t_1 +m_2^2 - m_1^2 + p^2 \pm 2\sqrt{p^2(m_2^2+t_2)}, \\
    z_1^{\pm} &= m_2^2 - m_1^2 + p^2 \pm 2\sqrt{p^2m_2^2} .
\end{align}

\subsection{One loop diagram for the heavy quark potential}\label{sec: 5.3}
\begin{figure}
    \centering
    \includegraphics[width=0.4\linewidth]{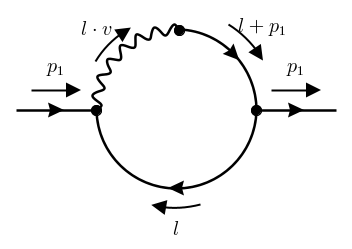}
    \caption{One-loop diagram for the heavy quark potential. The wavy line denotes a propagator for the static source}
    \label{fig:one loop heavy quark}
\end{figure}

As the previous examples focused on simple cases with standard quadratic propagators, we now provide a one-loop example that includes a linear propagator. Specifically, we consider the one-loop triangle diagram relevant to the heavy quark potential, which contains a propagator linear in the loop momentum, shown in figure \ref{fig:one loop heavy quark}. The corresponding general Feynman integral is

\begin{equation}
    I(a_1,a_2,a_3) = \int \frac{d^Dl}{(l^2)^{a_1} [(l+p_1)^2]^{a_2}(l\cdot v+i 0)^{a_3}}
    \rightarrow \int_{\Gamma} \frac{dz_1dz_2dz_3}{z_1^{a_1}z_2^{a_2}z_3^{a_3}}
    P(z_1,z_2,z_3)^{\frac{D}{2}-2},
\end{equation}
with $v\cdot p_1=0$. The Baikov variables are
\begin{equation}
  z_1=l^2,\quad z_2=(l+p_1)^2,\quad z_3=l\cdot v,
\end{equation}
and the Baikov polynomial is
\begin{align}
  P(z_1,z_2,z_3) = \begin{vmatrix}
    l^2 & l\cdotp p_1 & l\cdot v \\
    l\cdot p_1 & p_1^2 & 0 \\
    l\cdot v & 0 & v^2
    \end{vmatrix} =
    -\frac{1}{4} \Big[v^2
   (z_1-z_2+p_1^2)^2 - 4v^2p_1^2z_1  - 4p_1^2z_3^2,
    \Big].
\end{align}
This integral has two master integrals $I(1,1,1)$, $I(1,1,0)$, and the expansion formula is
\begin{align}
     \int_{\Gamma'}  &\frac{dz_1dz_2dz_3}{z_1z_2z_3}
    P(z_1+t_1,z_2+t_2,z_3+t_3)^{\frac{D}{2}-2} 
    = GF_3(\vec{t})\int_{\Gamma} \frac{dz_1dz_2dz_3}{z_1z_2z_3}
    P(z_1,z_2,z_3)^{\frac{D}{2}-2}  \nonumber\\
    & 
    + GF_{2,\hat{3}}(\vec{t}) \int_{\Gamma} \frac{dz_1dz_2dz_3}{z_1z_2}
    P(z_1,z_2,z_3)^{\frac{D}{2}-2} .
\end{align}
Taking the residue at $z_1=z_2=z_3=0$, we have
\begin{equation}
  P(t_1,t_2,t_3)^{\frac{D}{2}-2} = GF_3(\vec{t}) P(0,0,0)^{\frac{D}{2}-2}.
\end{equation}
Then, we take the residue at $z_1=z_2=0$, obtaining
\begin{small}
    \begin{equation}
  \int_{\tilde{z}_3^-(\vec{t})}^{\tilde{z}_3^+(\vec{t})} \frac{dz_3}{z_3} P(t_1, t_2, z_3+t_3)^{\frac{D}{2}-2}
  = GF_3(\vec{t}) \int_{z_3^-}^{z_3^+} \frac{dz_3}{z_3} P(0,0,z_3)^{\frac{D}{2}-2}
  + GF_{2,\hat{1}}(\vec{t}) \int_{z_3^-}^{z_3^+} dz_3P(0,0,z_3)^{\frac{D}{2}-2}.
\end{equation}
\end{small}
where \( \tilde{z}_3^{\pm}(\vec{t}) \) are the two roots of $P(t_1,t_2,z_3+t_3)=0$ ,
\( z_3^{\pm} \) are the two roots of \( P(0,0,z_3)= 0 \), 
\begin{small}
    \begin{align}
   \tilde{z}_3^{\pm} (\vec{t})&= (2p_1^2)^{-1}\Big[
   -2p_1^2t_3 \pm \sqrt{-p_1^6v^2+2p_1^4v^2t_1-p_1^2v^2t_1^2+2p_1^4v^2t_2+2p_1^2v^2t_1t_2-p_1^2v^2t_2^2}
   \Big] ,\\
   z_3^{\pm}& =  \pm \frac{\sqrt{-p_1^2v^2}}{2}.
\end{align}
\end{small}
The expression for $GF_{2,\hat{3}}(\vec{t})$ is
\begin{align}
  GF_{2,\hat{3}}(\vec{t}) &= 
  \Bigg(\frac{(\tilde{z}_3^+ (\vec{t})- \tilde{z}_3^-(\vec{t}))}{z_3^+-z_3^-} \Bigg)^{D-3}
  \frac{1}{\tilde{z}_3^-(\vec{t})}
  \sideset{_2}{_1}{\mathop{\mathrm{F}}}\left( \begin{array}{c} 1,\frac{D}{2}-1 \\ D-2\end{array} \middle| 1-\frac{\tilde{z}_3^+(\vec{t})}{\tilde{z}_3^-(\vec{t})}  \right) \nonumber\\
  &\qquad - \Bigg(\frac{P(t_1,t_2)}{P(0,0)}\Bigg)^{\frac{D}{2}-2}
  \frac{1}{z_3^-}
  \sideset{_2}{_1}{\mathop{\mathrm{F}}}\left( \begin{array}{c} 1,\frac{D}{2}-1 \\ D-2 \end{array} \middle| 1-\frac{z_3^+}{z_3^-}  \right).
\end{align}

\section{Higher loop examples}\label{sec: 6}

Since the Baikov representation is universally applicable to multi-loop integrals compared to the traditional momentum representation, we now present several higher-loop examples to demonstrate the application of our method in such cases.

\subsection{Two loop and three loop vacuum diagram}\label{sec: 6.1}

\begin{figure}
    \centering
    \includegraphics[width=0.7\linewidth]{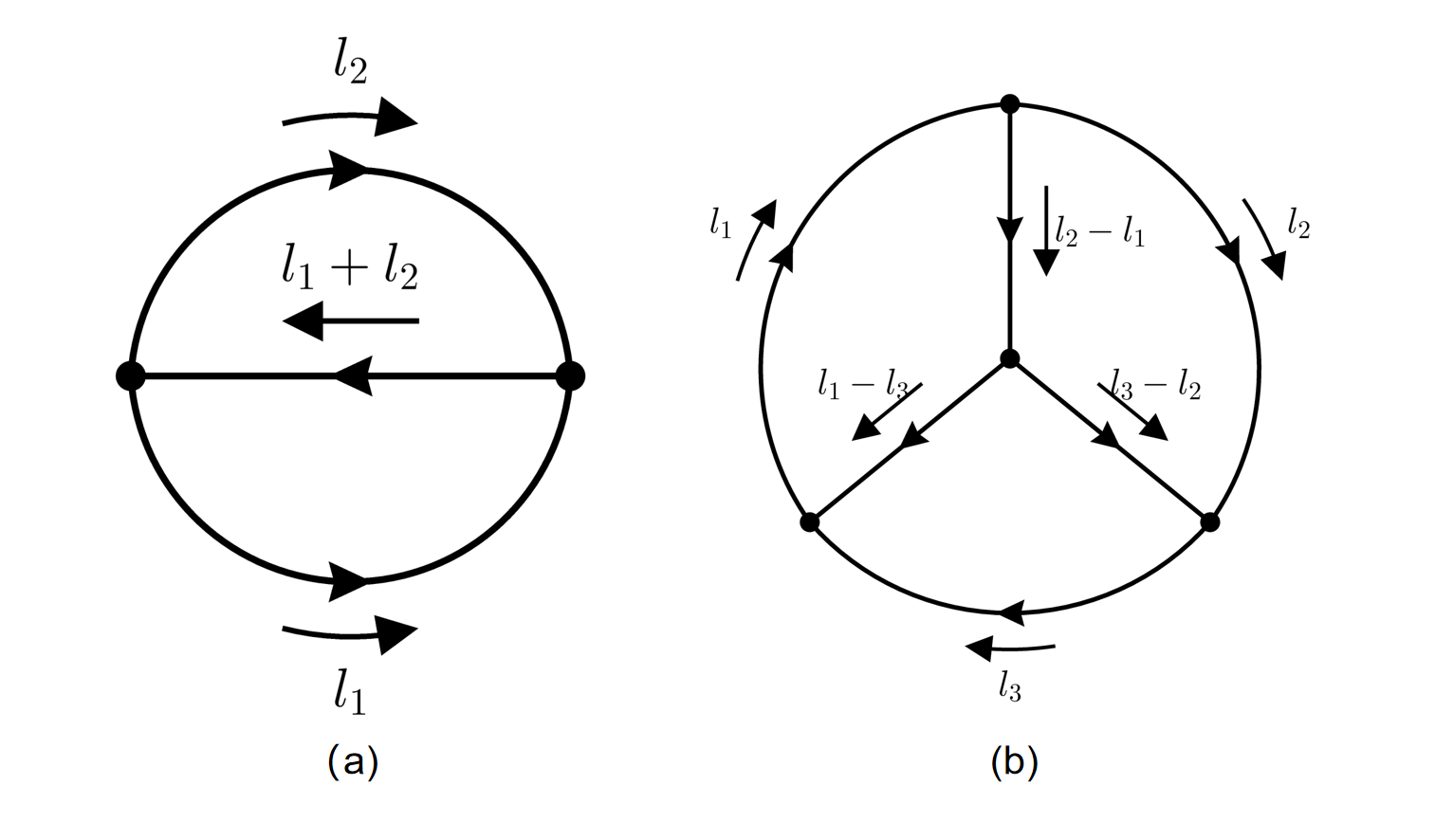}
    \caption{(a) Two loop vacuum diagram, (b) Three loop vacuum diagram}
    \label{fig:23loopvac}
\end{figure}
We first consider the two-loop and three-loop vacuum diagrams, as shown in the figure \ref{fig:23loopvac}. In this subsection, we consider examples where the three propagators in the two-loop vacuum diagram carry independent momenta, and, for simplicity, all six propagators in the three-loop vacuum diagram are assumed to have the same mass. Their Feynman integrals and the corresponding Baikov representations are as follows:
\begin{small}
    \begin{align}
         &I_{2-vac}(a_1,a_2,a_3) = \int \frac{d^Dl_1 d^Dl_2}{[l_1^2-m_1^2]^{a_1} [l_2^2-m_2^2]^{a_2}[(l_1+l_2)^2-m_3^2]^{a_3}}
  \rightarrow \int_{\Gamma} \frac{dz_1dz_2dz_3}{z_1^{a_1}z_2^{a_2}z_3^{a_3}}
  P_{2-vac}(z_1,z_2,z_3)^{\frac{D-3}{2}},\nonumber\\
  &I_{3-vac}(a_1,a_2,a_3,a_4,a_5,a_6) = \int \frac{d^Dl_1 d^Dl_2d^Dl_3}{[l_1^2-m^2]^{a_1} [l_2^2-m^2]^{a_2}[l_3^2-m^2]^{a_3}[(l_1-l_2)^2-m^2]^{a_4}[(l_2-l_3)^2-m^2]^{a_5}} \nonumber\\
  &\times \frac{1}{[(l_1-l_3)^2-m^2]^{a_6}}\rightarrow \int_{\Gamma} \frac{dz_1dz_2dz_3dz_4dz_5dz_6}{z_1^{a_1}z_2^{a_2}z_3^{a_3}z_4^{a_4}z_5^{a_5}z_6^{a_6}}
  P_{3-vac}(z_1,z_2,z_3,z_4,z_5,z_6)^{\frac{D}{2}-2}.
    \end{align}
\end{small}
The corresponding Baikov polynomials for these two diagrams are:
\begin{footnotesize}
    \begin{align}
         &P_{2-vac}(\vec{z}) = \begin{vmatrix}
    l_1^2 & l_1 \cdot l_2 \\
    l_1 \cdot l_2 & l_2^2
    \end{vmatrix} 
    = -\frac{1}{4} \Big[
    z_1^2 + z_2^2 + z_3^2 - 2 (z_1 z_2 + z_1 z_3 + z_2 z_3 ) 
    + 2( m_1^2 -  m_2^2 -  m_3^2) z_1\nonumber\\
& +2 (- m_1^2 +  m_2^2 -  m_3^2) z_2 + 2(- m_1^2 -  m_2^2 +  m_3^2) z_3 + m_1^4 + m_2^4+ m_3^4 - 2 (m_1^2 m_2^2 + m_1^2 m_3^2 + m_2^2 m_3^2 )
    \Big], \\
      &P_{3-vac}(\vec{z}) = \begin{vmatrix}
    l_1^2 & l_1 \cdot l_2 & l_1\cdot l_3\\
    l_1 \cdot l_2 & l_2^2 & l_2\cdot l_3 \\
    l_1\cdot l_3 & l_2\cdot l_3 & l_3^2
    \end{vmatrix}
    = \frac{1}{4} \Big[  2 m^6 + z_3^2 \left( -m^2 - z_4 \right) - m^2 z_4^2 + z_1^2 \left( -m^2 - z_5 \right) + z_2^2 \left( -m^2 - z_6 \right) \nonumber\\
    & - m^2 z_5^2  + m^4 z_6 - m^2 z_6^2 
    + z_5 \left( m^4 + m^2 z_6 \right) + z_4 \left( m^4 + z_5 \left( m^2 - z_6 \right) + m^2 z_6 \right) + z_2 \big( m^4 - z_6^2 + z_4 \left( m^2 + z_6 \right) \nonumber\\
    &+ z_5 \left( m^2 + z_6 \right) + z_3 \left( m^2 + z_4 - z_5 + z_6 \right) \big) 
    + z_3 \left( m^4 - z_4^2 + m^2 z_5 + m^2 z_6 + z_4 \left( z_5 + z_6 \right) \right) \nonumber\\
    &+ z_1 \left( m^4 - z_5^2 + z_4 \left( m^2 + z_5 \right) + z_3 \left( m^2 + z_4 + z_5 - z_6 \right) + m^2 z_6 + z_5 z_6 + z_2 \left( m^2 - z_4 + z_5 + z_6 \right) \right)
    \Big].
    \end{align}
\end{footnotesize}
It can be seen that both of these Baikov polynomials are quadratic in each $z_i$. Besides, the master integrals for the two-loop vacuum diagram are $I_{2-vac}(1,1,1)$, $I_{2-vac}(1,1,0)$, $I_{2-vac}(1,0,1)$, and $I_{2-vac}(0,1,1)$. For the three-loop vacuum diagram, the master integrals for the top and sub top sectors are
\begin{align}
     &I_{3-vac}(1,1,1,1,1,1),\ I_{3-vac}(1,1,1,1,1,0),\ I_{3-vac}(1,1,1,1,0,1),\ I_{3-vac}(1,1,1,0,1,1),\nonumber\\ &I_{3-vac}(1,1,0,1,1,1),\ I_{3-vac}(1,0,1,1,1,1),\ I_{3-vac}(0,1,1,1,1,1).
\end{align}
It is straightforward to see that the generating functions for the reduction coefficients of both the top and sub top sectors can be obtained using the method described earlier. Specifically, by taking the residues at zero of the corresponding Baikov variables, determining the two roots of the quadratic polynomial, and applying formula \eqref{eq:integral 1}, \eqref{eq:integral 2} and \eqref{eq:GF n to n-1}, the closed-form expressions can be constructed directly.
\begin{align}
    GF(\vec{t}) &= 
    \Bigg( \frac{\tilde{C}(\vec{t})}{C} \Bigg)^{\gamma}
    \Bigg(\frac{[\tilde{z}^+ (\vec{t})- \tilde{z}^-(\vec{t})]^2}{[z^+-z^-]^2} \Bigg)^{\gamma+\frac{1}{2}}
    \frac{1}{\tilde{z}^-(\vec{t})}
    \sideset{_2}{_1}{\mathop{\mathrm{F}}}\left( \begin{array}{c} 1,\gamma+1 \\ 2\gamma+2 \end{array} \middle| 1-\frac{\tilde{z}^+(\vec{t})}{\tilde{z}^-(\vec{t})}  \right) \nonumber\\
    &\qquad - \Bigg(\frac{P(\vec{t})}{P(\vec{0})}\Bigg)^{\gamma}
    \frac{1}{z^-}
    \sideset{_2}{_1}{\mathop{\mathrm{F}}}\left( \begin{array}{c} 1,\gamma+1 \\ 2\gamma+2 \end{array} \middle| 1-\frac{z^+}{z^-}  \right)
    \Bigg].
    \label{eq:GF of type II}
  \end{align}
For the other master integrals of the three-loop vacuum diagram, the orders of the poles associated with the propagators are either zero or one. In principle, their generating functions can also be computed iteratively. However, the integrations over the Baikov variables involved are highly nontrivial, making it difficult to obtain a closed-form expression.

\subsection{The massless sunset-type diagram with a vertical propagator}\label{sec: 6.2}

From the two-loop and three-loop vacuum diagrams discussed above, it can be seen that, in the Baikov representation, these examples share exactly the same structure as one-loop diagrams. This highlights one of the key advantages of the Baikov representation: its broader applicability.
\begin{figure}
    \centering
    \includegraphics[width=0.5\linewidth]{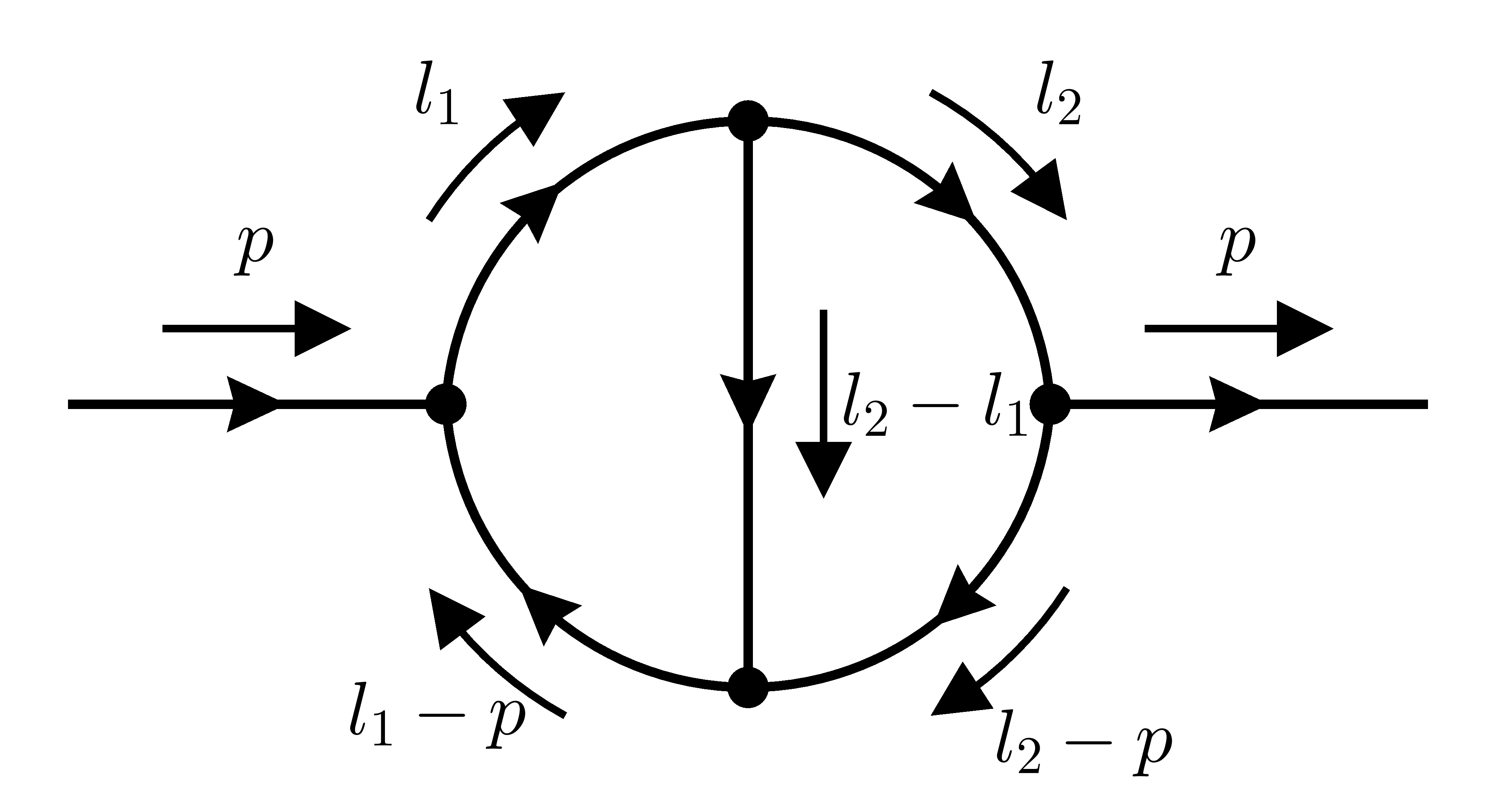}
    \caption{The massless sunset-type diagram with a vertical propagator}
    \label{fig:suncut 01}
\end{figure}
Next, we consider a slightly different example: a massless sunset-type diagram with a vertical propagator, as shown in the figure \ref{fig:suncut 01}. The corresponding Feynman integral is given by:

\begin{align}
    I(a_1,a_2,a_3,a_4,a_5) &= \int \frac{d^Dl_1 d^Dl_2}{(l_1^2)^{a_1} [(l_1-p)^2]^{a_2}(l_2^2)^{a_3}[(l_2-p)^2]^{a_4}[(l_1-l_2)^2]^{a_5}} \nonumber\\
  &\rightarrow \int_{\Gamma} \frac{dz_1dz_2dz_3dz_4dz_5}{z_1^{a_1}z_2^{a_2}z_3^{a_3}z_4^{a_4}z_5^{a_5}}
  P(z_1,z_2,z_3,z_4,z_5)^{\frac{D}{2}-2}.
\end{align}
The Baikov polynomial is
\begin{align}
  &P(z_1,z_2,z_3,z_4,z_5) = \begin{vmatrix}
    l_1^2 & l_1 \cdot l_2 & l_1 \cdot p \\
    l_1 \cdot l_2 & l_2^2 & l_2 \cdot p \\
    l_1 \cdot p & l_2 \cdot p & p^2
    \end{vmatrix} =
    -\frac{1}{4} \Big[
     z_2^2 z_3 + z_2 z_3^2 + z_1^2 z_4 + z_1 z_4^2 + s z_5^2 \nonumber\\
    & - z_1 z_2 z_3 - z_1 z_2 z_4 - z_1 z_3 z_4 - z_2 z_3 z_4 - z_2 z_3 z_5 - z_1 z_4 z_5 + z_1 z_3 z_5  + z_2 z_4 z_5 \nonumber\\
&+ s( z_1 z_2 -  z_2 z_3 -  z_1 z_4 +  z_3 z_4 -  z_1 z_5 -  z_2 z_5 -  z_3 z_5 -  z_4 z_5) + s^2 z_5 
    \Big],
\end{align}
where $s=p^2$.
This family of integrals has three master integrals $I(1,1,1,1,0)$, $I(1,0,0,1,1)$, $I(0,1,1,0,1)$. The generating function in the Baikov representation can be expanded in terms of these three master integrals, and is written as:
\begin{align}
     \int_{\Gamma'}  &\frac{dz_1dz_2dz_3dz_4dz_5}{z_1z_2z_3z_4z_5}
    P(z_1+t_1,z_2+t_2,z_3+t_3,z_4+t_4,z_5+t_5)^{\frac{D}{2}-2} \nonumber\\
    &= GF_{4,\hat{5}}(t_1,t_2,t_3,t_4,t_5) \int_{\Gamma} \frac{dz_1dz_2dz_3dz_4dz_5}{z_1z_2z_3z_4}
    P(z_1,z_2,z_3,z_4,z_5)^{\frac{D}{2}-2}  \nonumber\\
    &+ GF_{3,\widehat{23}}(t_1,t_2,t_3,t_4,t_5) \int_{\Gamma} \frac{dz_1dz_2dz_3dz_4dz_5}{z_1z_4z_5}
    P(z_1,z_2,z_3,z_4,z_5)^{\frac{D}{2}-2}  \nonumber\\
    &+ GF_{3,\widehat{14}}(t_1,t_2,t_3,t_4,t_5) \int_{\Gamma} \frac{dz_1dz_2dz_3dz_4dz_5}{z_2z_3z_5}
    P(z_1,z_2,z_3,z_4,z_5)^{\frac{D}{2}-2} .
\end{align}
Since the top sector of the master integrals involves four propagators, we take the corresponding residues at $z_1 = z_2 = z_3 = z_4 = 0$, and obtain
\begin{equation}
  \int_{\tilde{z}_5^-(\vec{t})}^{\tilde{z}_5^+(\vec{t})} dz_5 \frac{P(t_1,t_2,t_3,t_4,t_5+z_5)^{\frac{D}{2}-2}}{z_5}
  = GF_{4,\hat{5}}(\vec{t})
  \int_{z_5^-}^{z_5^+} dz_5 P(0,0,0,0,z_5)^{\frac{D}{2}-2},
\end{equation}
where $\tilde{z}_5^{\pm}(\vec{t})$ and $z_5^\pm$ are
\begin{align}
   \tilde{z}_5^{\pm}(\vec{t}) &= \frac{1}{2 s} \Big[-s^2+s \left(t_1+t_2+t_3+t_4-2 t_5\right)-\left(t_1-t_2\right) \left(t_3-t_4\right) \nonumber\\
  & \pm \sqrt{-2 t_1 \left(s+t_2\right)+\left(s-t_2\right){}^2+t_1^2} \sqrt{-2 t_3 \left(s+t_4\right)+\left(s-t_4\right){}^2+t_3^2} \Big],\\
   z_5^{+} &=0,\quad z_5^{-}=-s .
\end{align}
Similarly the expression for $GF_{4,\hat{5}}(\vec{t})$ is
\begin{equation}
    GF_{4,\hat{5}}(\vec{t}) = \Bigg(\frac{(\tilde{z}_5^+(\vec{t}) - \tilde{z}_5^-(\vec{t}))}{z_5^+-z_5^-} \Bigg)^{D-3}
     \frac{1}{\tilde{z}_5^-(\vec{t})}
     \sideset{_2}{_1}{\mathop{\mathrm{F}}}\left( \begin{array}{c} 1,\frac{D}{2}-1\\ D-2 \end{array} \middle| 1-\frac{\tilde{z}_5^+(\vec{t})}{\tilde{z}_5^-(\vec{t})}  \right) .
     \label{eq:GF of Suncut}
\end{equation}

Due to the condition $z_5^+ = 0$, the change of variables introduced in \eqref{eq:trans 1} is no longer applicable in the present case. To extract the reduction coefficients from the generating function, we instead adopt the alternative transformation in \eqref{eq:trans 3}, which allows us to rewrite the relevant expressions in terms of elementary functions without involving hypergeometric functions. This transformation will be described in detail in the next subsection, where we demonstrate how the generating function yields the desired reduction coefficients in closed analytic form.

For master integrals corresponding to lower topologies such as $I(0,1,1,0,1)$ the residue method can still be applied to construct generating functions, although the resulting integrals generally involve more variables and increased analytic complexity. In certain cases, however, partial parameterization may allow for closed-form expressions. For instance, by introducing deformation parameters $t_1$ and $t_4$ while keeping the powers of the remaining propagators fixed as $a_2 = a_3 = a_5 = 1$, the generating function for $I(0,1,1,0,1)$ can be written in terms of the Appell $F_2$ function. First, we write down the expansion: 

\begin{equation}
\begin{aligned}
& \int_{\Gamma^{\prime}} \frac{d z_1 d z_2 d z_3 d z_4 d z_5}{z_1 z_2 z_3 z_4 z_5} P\left(z_1+t_1, z_2, z_3, z_4+t_4, z_5\right)^{\frac{D}{2}-2}\\
=& G F_{4, \hat{5}}\left(t_1, t_4\right) \int_{\Gamma} \frac{d z_1 d z_2 d z_3 d z_4 d z_5}{z_1 z_2 z_3 z_4} P\left(z_1, z_2, z_3, z_4, z_5\right)^{\frac{D}{2}-2} \\
+& G F_{3, \widehat{23}}\left(t_1, t_4\right) \int_{\Gamma} \frac{d z_1 d z_2 d z_3 d z_4 d z_5}{z_1 z_4 z_5} P\left(z_1, z_2, z_3, z_4, z_5\right)^{\frac{D}{2}-2}\\
+&G F_{3, \widehat{14}}\left(t_1, t_4\right) \int_{\Gamma} \frac{d z_1 d z_2 d z_3 d z_4 d z_5}{z_2 z_3 z_5} P\left(z_1, z_2, z_3, z_4, z_5\right)^{\frac{D}{2}-2}.
\end{aligned}
\end{equation}
Taking the residues of both sides at $z_2 = z_3 = z_5 = 0$, we obtain:

\begin{equation}
\int_{\Gamma^{\prime}} \frac{d z_1 d z_4}{z_1 z_4} P\left(z_1+t_1, 0,0, z_4+t_4, 0\right)^{\frac{D}{2}-2}=G F_{3, \widehat{14}}\left(t_1, t_4\right) \int_{\Gamma} d z_1 d z_4 P\left(z_1, 0,0, z_4, 0\right)^{\frac{D}{2}-2}.
\end{equation}
The integration regions and corresponding results for the two double integrals are as follows:
\begin{equation}
\begin{aligned}
& \int_{-t_1}^{\infty} d z_1 \int_{-t_4}^{\infty} d z_4 \frac{1}{z_1 z_4} P\left(z_1+t_1, 0,0, z_4+t_4, 0\right)^{\frac{D}{2}-2}\\
=&\int_0^{\infty} d z_1 \int_0^{\infty} d z_4 \frac{1}{\left(z_1-t_1\right)\left(z_4-t_4\right)}\left[-\frac{1}{4} z_1 z_4\left(-s+z_1+z_4\right)\right]^{\frac{D}{2}-2} \\
=&\left(-\frac{1}{4}\right)^{\frac{D}{2}-2}(-s)^{\frac{3 D}{2}-6} \frac{\Gamma\left(\frac{D}{2}-2\right)^2 \Gamma\left(\frac{3 D}{2}-6\right)}{\Gamma\left(2-\frac{D}{2}\right)} F_2\left(6-\frac{3 D}{2}, 1,1,3-\frac{D}{2}, 3-\frac{D}{2}, \frac{t_1}{s}, \frac{t_4}{s}\right),
\end{aligned}
\end{equation}
and
\begin{equation}
\int_0^{\infty} d z_1 \int_0^{\infty} d z_4\left[-\frac{1}{4} z_1 z_4\left(-s+z_1+z_4\right)\right]^{\frac{D}{2}-2}=\left(-\frac{1}{4}\right)^{\frac{D}{2}-2}(-s)^{\frac{3 D}{2}-4} \frac{\Gamma\left(\frac{D}{2}-1\right)^2 \Gamma\left(4-\frac{3 D}{2}\right)}{\Gamma\left(2-\frac{D}{2}\right)}.
\end{equation}
Here, $F_2$ denotes the Appell hypergeometric function of the second kind, defined by
\begin{equation}
F_2(a; b_1, b_2; c_1, c_2; x, y) = \sum_{m=0}^\infty \sum_{n=0}^\infty \frac{(a)_{m+n} \, (b_1)_m \, (b_2)_n}{(c_1)_m \, (c_2)_n \, m! \, n!} \, x^m y^n,
\end{equation}
Thus we obtain the generating function $GF_{3,\widehat{14}}(t_1,t_4)$ is

\begin{equation}
G F_{3, \widehat{14}}\left(t_1, t_4\right)=\frac{\left(\frac{3 D}{2}-4\right)\left(\frac{3 D}{2}-5\right)}{s^2\left(\frac{D}{2}-2\right)^2} F_2\left(6-\frac{3 D}{2}, 1,1,3-\frac{D}{2}, 3-\frac{D}{2}, \frac{t_1}{s}, \frac{t_4}{s}\right).
\end{equation}
In this example, although we do not introduce the auxiliary parameters $t_2$, $t_3$, and $t_5$, the use of $t_1$ and $t_4$ alone already leads to a significant simplification in the computation of the reduction coefficients, compared to performing the reduction without the generating function approach.


\subsection{The sunset-type diagram with four propagators}\label{sec: 6.3}

\begin{figure}
    \centering
    \includegraphics[width=0.4\linewidth]{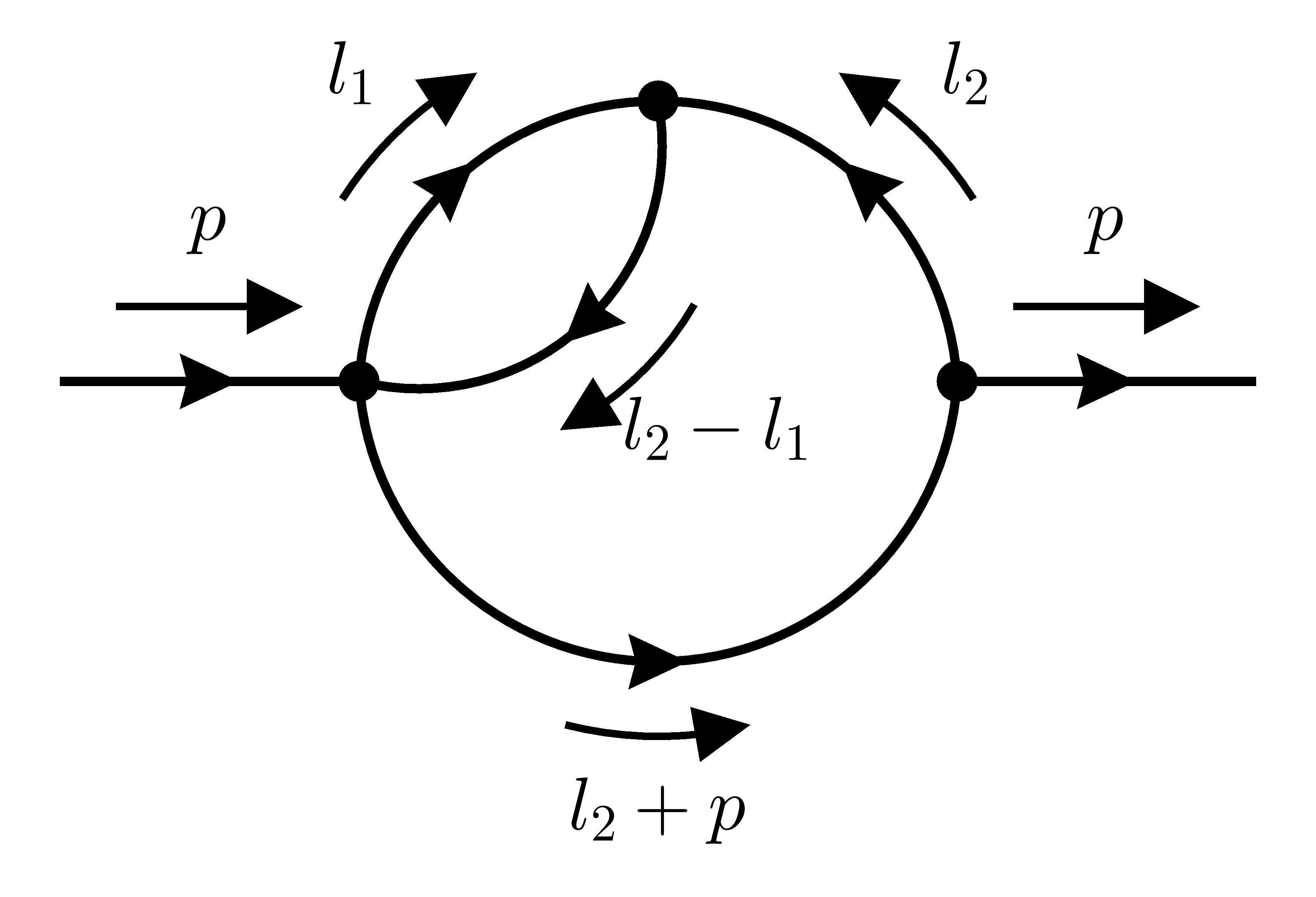}
    \caption{The sunset-type diagram with four propagators}
    \label{Fig:suncut with 4 prop}
\end{figure}

As seen from all the previous examples, they satisfy the condition $N = L(L+1)/2 + LE$, meaning that all irreducible scalar products involving loop momenta can be linearly expressed in terms of the propagators. We now consider an example with fewer propagators: the sunset-type diagram discussed in the previous subsection, but with one propagator removed, as shown in the figure. The corresponding Feynman integral is given by:
\begin{align}
  I(a_1,a_2,a_3,a_4)& = \int \frac{d^Dl_1 d^Dl_2}{[l_1^2-m^2]^{a_1} [l_2^2-m^2]^{a_2}[(l_2+p)^2-m^2]^{a_3}[(l_1+l_2)^2-m^2]^{a_4}} \nonumber.
\end{align}
In this example, there are five scalar products involving loop momenta, namely $l_1^2$, $l_2^2$, $l_1 \cdot p$, $l_2 \cdot p$, and $l_1 \cdot l_2$. However, there are only four Baikov variables, given by:
\begin{equation}
  z_1=l_1^2-m^2,\quad z_2=l_2^2-m^2,\quad z_3=(l_2+p)^2-m^2,\quad  
  z_4=(l_1+l_2)^2-m^2.
\end{equation}
Therefore, we introduce an additional Baikov variable, $z_5 = l_1 \cdot p$, to complete the set. Then the baikov representation of this Feynman integral is 
\begin{align}
  I(a_1,a_2,a_3,a_4)\rightarrow \int_{\Gamma} \frac{dz_1dz_2dz_3dz_4dz_5}{z_1^{a_1}z_2^{a_2}z_3^{a_3}z_4^{a_4}}
  P(z_1,z_2,z_3,z_4,z_5)^{\frac{D}{2}-2}.
\end{align}
The Baikov polynomial is
\begin{footnotesize}
    \begin{align}
  &P(z_1,z_2,z_3,z_4,z_5) = \begin{vmatrix}
    l_1^2 & l_1 \cdot l_2 & l_1 \cdot p \\
    l_1 \cdot l_2 & l_2^2 & l_2 \cdot p \\
    l_1 \cdot p & l_2 \cdot p & p^2
    \end{vmatrix} = -\frac{1}{4} \Big[
   m^2 (z_2^2 +  z_3^2 + 4  z_5^2 )
   + s( z_1^2 +  z_2^2 +  z_4^2)+ z_1 z_2^2 + z_1 z_3^2 \nonumber\\
 &   + 4 z_2 z_5^2 - 2 z_2^2 z_5 
 -2 z_1 z_2 z_3 - 2 z_1 z_2 z_5 + 2 z_1 z_3 z_5 + 2 z_2 z_3 z_5 + 2 z_2 z_4 z_5 - 2 z_3 z_4 z_5 
-2 m^2( z_2 z_3 + z_2 z_5 - z_3 z_5 ) \nonumber\\
&- 2 s (z_1 z_3 + z_1 z_4 + z_2 z_4 + z_1 z_5 + z_2 z_5 - z_4 z_5 ) 
 -2 m^2 s (z_1 + z_3 +z_4+ z_5 )+ s^2 z_1 
-3 m^4 s + m^2 s^2
     \Big],
\end{align}
\end{footnotesize}
where $p^2=s$.
This family of Feynman integrals has only one top sector master integrals $I(1,1,1,1)$, and the expansion formula is
\begin{align}
     \int_{\Gamma'}  &\frac{dz_1dz_2dz_3dz_4dz_5}{z_1z_2z_3z_4}
    P(z_1+t_1,z_2+t_2,z_3+t_3,z_4+t_4,z_5)^{\frac{D}{2}-2} \nonumber\\
    &= GF_{4}(t_1,t_2,t_3,t_4) \int_{\Gamma} \frac{dz_1dz_2dz_3dz_4dz_5}{z_1z_2z_3z_4}
    P(z_1,z_2,z_3,z_4,z_5)^{\frac{D}{2}-2} +\dots.
\end{align}
Taking the residue at $z_1=z_2=z_3=z_4=0$, we have
\begin{equation}
  \int_{\tilde{z}_5^-(\vec{t})}^{\tilde{z}_5^+(\vec{t})} dz_5 P(t_1,t_2,t_3,t_4,z_5)^{\frac{D}{2}-2}
  = GF_{4}(t_1,t_2,t_3,t_4) 
  \int_{z_5^-}^{z_5^+} dz_5 P(0,0,0,0,z_5)^{\frac{D}{2}-2},
\end{equation}
where $ \tilde{z}_5^{\pm}(\vec{t}) $ are the two roots of $P(t_1,t_2,t_3,t_4,z_5)=0$ \footnote{Note that in this case, $\tilde{z}_5^{\pm}(\vec{t})$ are functions of $t_1, t_2, t_3,$ and $t_4$ only. Unlike the previous cases, $\tilde{z}_5^{\pm}(\vec{t})$ are the two roots of the equation $P(t_1, t_2, t_3, t_4, z_5) = 0$ rather than the two roots of $P(t_1, t_2, t_3, t_4, z_5 + t_5) = 0$.},
\( z_5^{\pm} \) are the two roots of \( P(0,0,0,0,z_5)= 0 \), 
\begin{align}
   &\tilde{z}_2^{\pm}(\vec{t}) = \frac{1}{4 \left(m^2+t_2\right)}
 \Big[m^2 \left(s+t_2-t_3\right)
 +\left(t_1+t_2-t_4\right) \left(s+t_2-t_3\right) \nonumber\\
& \pm \sqrt{2 t_1 \left(m^2+t_2+t_4\right)+2 t_2 \left(m^2+t_4\right)+\left(3 m^2-t_4\right) \left(m^2+t_4\right)-t_1^2-t_2^2} \nonumber\\
&\times \sqrt{s \left(4 m^2-s\right)+2 t_2 \left(s+t_3\right)+2 s t_3-t_2^2-t_3^2}\Big],\\
   &z_2^{\pm} = \frac{1}{4} \left(s\pm\sqrt{3} \sqrt{4 m^2 s-s^2}\right).
\end{align}
Finally the expression for $GF_{4}(t_1,t_2,t_3,t_4)$ is
\begin{align}
 GF_{4}(t_1,t_2,t_3,t_4) &=  \Bigg( \frac{m^2+t_2}{m^2} \Bigg)^{\frac{D}{2}-2}
  \Bigg( \frac{\tilde{z}_2^+(\vec{t})- \tilde{z}_2^-(\vec{t})}{z_2^+-z_2^-} \Bigg) ^{D-3}.
  \label{eq:GF of suncut with 4 prop}
\end{align}
Hence, we obtain the closed-form generating function for the reduction to the top sector master integral.

\subsection{Other examples}\label{sec: 6.4}

\begin{figure}
    \centering
    \includegraphics[width=1\linewidth]{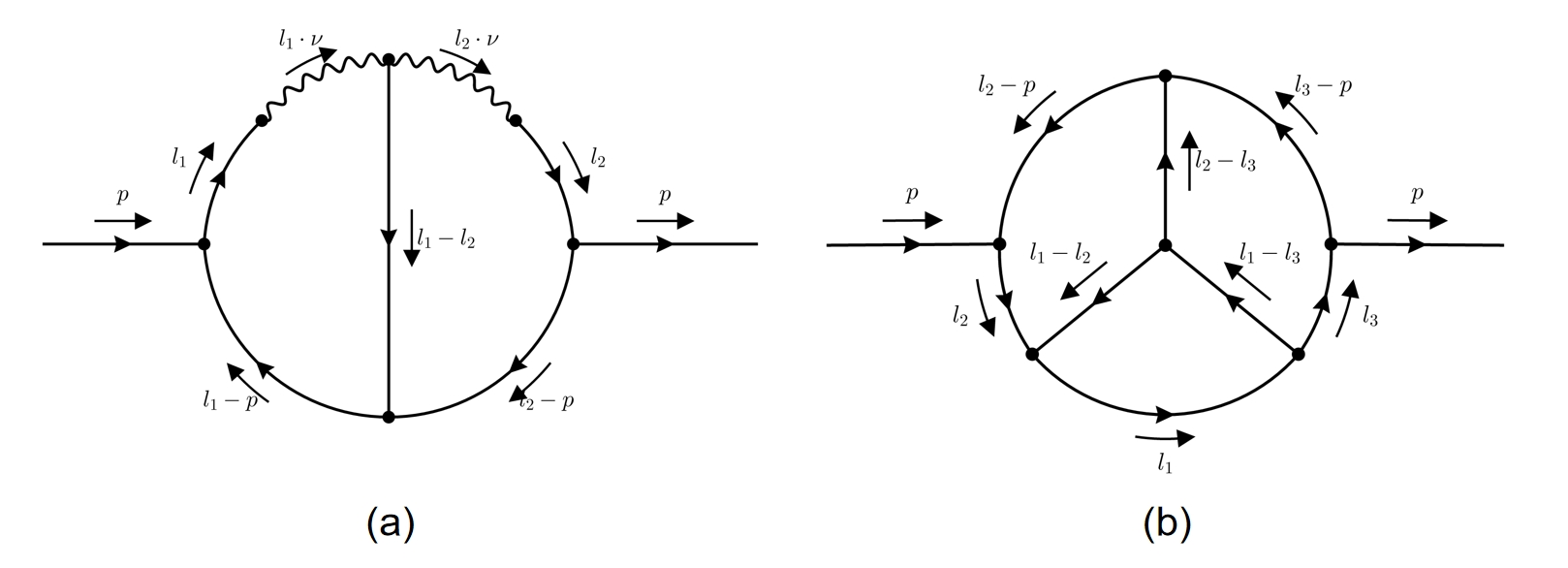}
    \caption{(a) Feynman diagrams corresponding to case A and case B. Wavy lines denote propagators for the static source, (b) Three loop One external momentum diagram.}
    \label{fig:other example}
\end{figure}
At the end of this section, we provide two additional examples to illustrate the applicability of the method proposed in this paper. The first example is the two-loop Feynman integrals for the heavy quark potential, corresponding to figure \ref{fig:other example} (a). The second example is a three-loop self-energy diagram involving three loop momenta and one external momentum, as shown in figure \ref{fig:other example} (b). Their Feynman integrals and corresponding Baikov forms are
\begin{small}
    \begin{align}
I_1(a_1,a_2,a_3,a_4,a_5,a_6,a_7) &= \int \frac{d^Dl_1 d^Dl_2}{(l_1^2)^{a_1} (l_2^2)^{a_2}[(l_1-p)^2]^{a_3}[(l_2-p)^2]^{a_4}[(l_1-l_2)^2]^{a_5}(l_1\cdot v)^{a_6}(l_2\cdot v)^{a_7}} \nonumber\\
  &\rightarrow \int_{\Gamma} \frac{dz_1dz_2dz_3dz_4dz_5dz_6dz_7}{z_1^{a_1}z_2^{a_2}z_3^{a_3}z_4^{a_4}z_5^{a_5}z_6^{a_6}z_7^{a_7}}
  P_1(z_1,z_2,z_3,z_4,z_5,z_6,z_7)^{\frac{D-5}{2}},
\end{align}
\end{small}
with $v\cdot p=0$, and
\begin{small}
    \begin{align}
I_2(a_1,a_2,a_3,&a_4,a_5,a_6,a_7,a_8) = \int \frac{d^Dl_1 d^Dl_2d^Dl_3}{[l_1^2-m^2]^{a_1} [l_2^2-m^2]^{a_2} [l_3^2-m^2]^{a_3}[(l_2-p)^2-m^2]^{a_4}  }\nonumber\\
\times & \frac{1}{  [(l_3-p)^2-m^2]^{a_5} [(l_1-l_2)^2-m^2]^{a_6}  [(l_1-l_3)^2-m^2]^{a_7}
  [(l_2-l_3)^2-m^2]^{a_8}} 
  \nonumber\\
  & \rightarrow \int_{\Gamma} \frac{dz_1dz_2dz_3dz_4dz_5dz_6dz_7dz_8dz_9}{z_1^{a_1}z_2^{a_2}z_3^{a_3}z_4^{a_4}z_5^{a_5}z_6^{a_6}z_7^{a_7}z_8^{a_8}}
  P_2(z_1,z_2,z_3,z_4,z_5,z_6,z_7,z_8,z_9)^{\frac{D-5}{2}}.
\end{align}
\end{small}
It is important to note that, in the first integral, there are seven scalar products involving loop momenta. This number matches the number of propagators, and thus the number of Baikov variables. However, in the second integral, there are nine scalar products involving loop momenta, but only eight propagators. Therefore, it is necessary to introduce an additional auxiliary Baikov variable, denoted by $z_9 = l_1\cdot p$, which is similar to the example given in the previous subsection. 

The top sector master integral for each of these two integral families is unique, given by $I_1(1,1,1,1,0,1,1)$ and $I_2(1,1,1,1,1,1,1,1)$, respectively. To compute the generating functions for the reduction coefficients onto the top sector master integrals, we need take the corresponding residues at $z_1 = z_2 =z_3=z_4=z_6=z_7 =0$ and $z_1=\cdots =z_8 = 0$, respectively. 
As a result of taking the residues at $z_1 = z_2 = 0$ for $I_1(1,1,1,1,1)$ and at $z_3 = 0$ for $I_2(1,1,1,1,1,1)$, the remaining integrals reduce to single definite integrals over $z_5$ and $z_9$, respectively. Note that in both cases, the corresponding Baikov polynomials are quadratic functions of $z_5$ or $z_9$. Therefore, the computation of the generating functions for the reduction coefficients can be carried out following the general procedure outlined in section \ref{sec: 6.2} and \ref{sec: 6.3} respectively. Specifically, by determining the two roots of the quadratic Baikov polynomials and applying the integral formulas presented earlier, the closed-form expressions for these generating functions can be explicitly obtained.

From the examples presented in this section, it becomes clear that, when working in the Baikov representation, the computation of generating functions reveals an underlying structural similarity between examples that appear entirely different in the traditional Feynman parametrization. For instance, the two-loop and three-loop vacuum diagrams exhibit nearly identical structures to that of the one-loop diagram. Similarly, the two examples discussed here share an almost identical structure with the examples provided in the previous subsection.

In summary, as long as the following characteristics are satisfied, the method and formulas developed in this paper can be applied.

\begin{itemize}
    \item(A) The number of propagators in the family of Feynman integrals is equal to, or at most one less than, the number of independent scalar products involving loop momenta. And The number of propagators in the corresponding top and sub top sector master integrals is equal to, or at most one less than, the number of independent scalar products involving both loop and external momenta. These two conditions ensure that the remaining definite integrals to be evaluated are single-variable integrals, which makes it possible to obtain closed-form expressions for the generating functions.

    \item(B) The number of independent equations for the generating functions of the reduction coefficients, obtained by taking residues with respect to different Baikov variables, is equal to the number of corresponding master integrals. This condition ensures that there are sufficiently many equations to uniquely determine the generating functions as unknowns.

    \item (C) The Baikov polynomial is quadratic with respect to the integration Baikov variable.

\end{itemize}

A comment on the three features listed above is in order: among them, the most important is condition~(B). This condition guarantees that there are sufficiently many independent equations to uniquely determine the generating functions. In contrast, conditions~(A) and~(C) are not strictly necessary. From a methodological perspective, the absence of either condition mainly results in a more non-trivial integration process, making it difficult to obtain a closed-form expression for the generating functions. Nevertheless, in principle, it is still possible to derive series solutions, although they may be more complicated. This also highlights the main challenge of directly computing the reduction coefficients via definite integrals in the Baikov representation without introducing generating functions.

\section{From Generating Function to Reduction coefficients}\label{sec: 7}

In this section, we discuss how to extract reduction coefficients of specific orders from the generating functions presented earlier. For generating functions of the following form:
\begin{equation}
GF(\vec{t})=\left(\frac{P(\vec{t})}{P(\vec{0})}\right)^{\gamma},
\end{equation}
it is straightforward to obtain the reduction coefficient by taking the corresponding derivative at $\vec{t} = \vec{0}$. Therefore, we focus on more general types of generating functions. Two main issues must be addressed:

\begin{enumerate}
  \item \textbf{Irrational terms in the generating function.} \\
  As seen in the previous sections, the limits of integration in the Baikov representation are often given by the roots of a quadratic polynomial, which introduces irrational expressions into the generating function. However, the reduction coefficients themselves are rational. A key task is to systematically eliminate these irrational terms to extract the correct rational coefficients.

  \item \textbf{Potential divergences in numerical evaluation.} \\
  Another issue, previously noted, is the potential divergence of the generating function under certain kinematic conditions. In particular, when the generating function involves hypergeometric functions whose variable depend on external momenta and masses, naive differentiation may lead to divergent in numerical evaluation and produce incorrect results.
\end{enumerate}

A careful analytical treatment of both issues is necessary in order to correctly and reliably extract the reduction coefficients from the generating functions. In the following, we analyze three types of generating functions that appear in this paper and outline how each type can be handled.

\subsection{Type I}\label{sec: 7.1}
Let us begin with the simplest type, this type of generating functions \red{has} the following form:
\begin{equation}
GF(\vec{t})=\left(\frac{C'(\vec{t})}{C}\right)^\gamma \left(\frac{\tilde{z}_i^+(\vec{t})-\tilde{z}_i^-(\vec{t})}{z_i^+-z_i^-}\right)^{\gamma'}.
\end{equation}
An example of this type appears in Section \ref{sec: 6.3} and again in the example of figure \ref{fig:other example} (b) in Section \ref{sec: 6.4}. As can be seen from these cases, the integrals that give rise to such generating functions typically satisfy the following conditions: (1) The number of propagators is one less than the number of scalar products involving loop momenta; (2) The number of propagators equals the number of the top sector master integral’s topology.

For generating functions of this type, it is sufficient to perform the following transformation:
\begin{equation}
GF(\vec{t})=\left(\frac{C'(\vec{t})}{C}\right)^\gamma \left(\frac{(\tilde{z}_i^+(\vec{t})-\tilde{z}_i^-(\vec{t}))^2}{(z_i^+-z_i^-)^2}\right)^{\gamma'/2}.
\end{equation}
Note that at $\vec{t} = \vec{0}$, we have $C'(\vec{t}) = C$ and $\tilde{z}_i^\pm(\vec{t}) = z_i^\pm$, so the structure of the generating function remains entirely rational after differentiation. This makes the use of this form particularly natural for extracting rational reduction coefficients.

\subsection{Type II}\label{sec: 7.2}
The second type of generating function appears in cases such as the one-loop example, the two-loop vacuum diagram with three propagators, and the three-loop vacuum diagram with six propagators discussed earlier in this paper. This type is characterized by the fact that (1) the number of propagators equals the number of scalar products involving loop momenta, and (2) both the top and sub top topologies are master integrals. For this reason, we refer to these as “one-loop-like” Feynman integrals. The generating function corresponding to the sub top sector in such cases typically takes the following form: 
\begin{align}
    GF(\vec{t}) &= 
    \Bigg( \frac{\tilde{C}(\vec{t})}{C} \Bigg)^{\gamma}
    \Bigg(\frac{[\tilde{z}^+ (\vec{t})- \tilde{z}^-(\vec{t})]^2}{[z^+-z^-]^2} \Bigg)^{\gamma+\frac{1}{2}}
    \frac{1}{\tilde{z}^-(\vec{t})}
    \sideset{_2}{_1}{\mathop{\mathrm{F}}}\left( \begin{array}{c} 1,\gamma+1 \\ 2\gamma+2 \end{array} \middle| 1-\frac{\tilde{z}^+(\vec{t})}{\tilde{z}^-(\vec{t})}  \right) \nonumber\\
    &\qquad - \Bigg(\frac{P(\vec{t})}{P(\vec{0})}\Bigg)^{\gamma}
    \frac{1}{z^-}
    \sideset{_2}{_1}{\mathop{\mathrm{F}}}\left( \begin{array}{c} 1,\gamma+1 \\ 2\gamma+2 \end{array} \middle| 1-\frac{z^+}{z^-}  \right).
  \end{align}
In order to extract the reduction coefficients via differentiation, we first apply the transformation given by
  \begin{equation}
    \begin{aligned}
    & \sideset{_2}{_1}{\mathop{\mathrm{F}}}\left(\begin{array}{c}
    1 , \gamma +1 \\
    2\gamma +2
    \end{array} \Bigg | 1-\frac{z^{+}}{z^{-}}\right)=
    \Bigg( \frac{z^{+}}{z^{-}} \Bigg)^{-\gamma-1}
    \sideset{_2}{_1}{\mathop{\mathrm{F}}}\left(\begin{array}{c}
    2\gamma +1 ,\gamma + 1\\
    2\gamma +2
    \end{array} \Bigg | 1-\frac{z^{-}}{z^{+}}\right),
    \end{aligned}
    \end{equation}
which brings the generating function into the following form:
    \begin{align}
        GF(\vec{t})= & 
        \Bigg( \frac{\tilde{C}(\vec{t})}{C} \Bigg)^{\gamma}
        \Bigg(\frac{[\tilde{z}^+ (\vec{t})- \tilde{z}^-(\vec{t})]^2}{[z^+-z^-]^2} \Bigg)^{\gamma+\frac{1}{2}}
        \frac{1}{\tilde{z}^-(\vec{t})}
        \Bigg( \frac{\tilde{z}^{+}(\vec{t})}{\tilde{z}^{-}(\vec{t})} \Bigg)^{-\gamma-1}
    \sideset{_2}{_1}{\mathop{\mathrm{F}}}\left(\begin{array}{c}
    2\gamma +1 ,\gamma + 1\\2\gamma +2
    \end{array} \Bigg | 1-\frac{\tilde{z}^{-}(\vec{t})}{\tilde{z}^{+}(\vec{t})}\right)
        \nonumber\\
        &\qquad - \Bigg(\frac{P(\vec{t})}{P(\vec{0})}\Bigg)^{\gamma}
        \frac{1}{z^-}
        \Bigg( \frac{z^{+}}{z^{-}} \Bigg)^{-\gamma-1}
        \sideset{_2}{_1}{\mathop{\mathrm{F}}}\left(\begin{array}{c}
        2\gamma +1 ,\gamma + 1\\2\gamma +2
        \end{array} \Bigg | 1-\frac{z^{-}}{z^{+}}\right).
      \end{align}
The main advantage of this transformation is that the hypergeometric function retains its form under differentiation:
      \begin{equation}
        \frac{d}{dz}  \sideset{_2}{_1}{\mathop{\mathrm{F}}}\left(\begin{array}{c}
        2\gamma +1 ,\gamma + 1\\  2\gamma +2
        \end{array} \Bigg | z\right)
        = \frac{2\gamma+1}{z}\left\{(1-z)^{-1-\gamma}- \sideset{_2}{_1}{\mathop{\mathrm{F}}}\left(\begin{array}{c}
        2\gamma +1 ,\gamma + 1\\  2\gamma +2
        \end{array} \Bigg | z\right)\right\} .
    \end{equation}
Consequently, after taking the appropriate-order derivatives with respect to the parameters $t_i$ and evaluating at $t_i = 0$, the reduction coefficient takes the following form:
    \begin{equation}
        H_1(\vec{a}) \cdot \sideset{_2}{_1}{\mathop{\mathrm{F}}}\left(\begin{array}{c}
            2\gamma +1 ,\gamma + 1\\  2\gamma +2
            \end{array} \Bigg | 1-\frac{z^{-}}{z^{+}}\right)
        + H_0(\vec{a}).
    \end{equation}
Here, $H_1(\vec{a})$ and $H_0(\vec{a})$ are elementary functions, although they are not necessarily rational functions in explicit form. In the following, we present a concrete example to illustrate that, after performing the necessary analytic simplifications, the function $H_1(\vec{a})$ vanishes, and the irrational parts of $H_0(\vec{a})$ cancel out. As a result, the final expression is purely rational, and $H_0(\vec{a})$ coincides exactly with the desired reduction coefficient. We consider the two-loop vacuum diagram with three propagators and focus on the master integral $I(1,1,0)$. The corresponding generating function has been given in Section \ref{sec: 6.1} and takes the following form:
    \begin{align}
        GF_{2,\hat{3}}(\vec{t}) =& \Bigg(\frac{[\tilde{z}^+_3 (\vec{t})- \tilde{z}^-_3(\vec{t})]^2}{[z^+_3-z^-_3]^2} \Bigg)^{\frac{D}{2}-1}
        \frac{1}{\tilde{z}^-_3(\vec{t})}
        \Bigg( \frac{z^{+}_3(\vec{t})}{z^{-}_3(\vec{t})} \Bigg)^{\frac{1-D}{2}}
        \sideset{_2}{_1}{\mathop{\mathrm{F}}}\left(\begin{array}{c}
        D-2 ,\frac{D-1}{2}\\D-1
        \end{array} \Bigg | 1-\frac{\tilde{z}^{-}_3(\vec{t})}{\tilde{z}^{+}_3(\vec{t})}\right)  \nonumber\\
        &\qquad - \Bigg(\frac{P(\vec{t})}{P(\vec{0})}\Bigg)^{\frac{D-3}{2}}
        \frac{1}{z_3^-}
        \Bigg( \frac{z^{+}_3}{z^{-}_3} \Bigg)^{\frac{1-D}{2}}
        \sideset{_2}{_1}{\mathop{\mathrm{F}}}\left(\begin{array}{c}
        D-2 ,\frac{D-1}{2}\\D-1
        \end{array} \Bigg | 1-\frac{z^{-}_3}{z^{+}_3}\right),
      \end{align}
where $\gamma=\frac{D-3}{2}$, and
      \begin{align}
        P(\vec{t}) 
          = &-\frac{1}{4} \Big[
          t_1^2 + t_2^2 + t_3^2 - 2 (t_1 t_2 + t_1 t_3 + t_2 t_3 ) 
          + 2( m_1^2 -  m_2^2 -  m_3^2) t_1
          +2 (- m_1^2 +  m_2^2 -  m_3^2) t_2\nonumber\\
      + & 2(- m_1^2 -  m_2^2 +  m_3^2) t_3 + m_1^4 + m_2^4+ m_3^4 - 2 (m_1^2 m_2^2 + m_1^2 m_3^2 + m_2^2 m_3^2 )
          \Big],\\
          \tilde{z}^{\pm}_3(\vec{t}) &= m_1^2+m_2^2-m_3^2+t_1+t_2-t_3 \pm 2\sqrt{(m_1^2+t_1)(m_2^2+t_2)},\\
        z^{\pm}_3 &= m_1^2+m_2^2-m_3^2 \pm 2\sqrt{m_1^2m_2^2},
        \\
         \tilde{C}(\vec{t})&=C=-\frac{1}{4}.
      \end{align}
Then the reduction coefficients are:
\begin{small}
     \begin{align}
        C_{2,\hat{3}}(a_1,a_2,a_3)& = \frac{1}{(a_1-1)!(a_2-1)!(a_3-1)!} 
        \Bigg(\frac{\partial}{\partial t_1}\Bigg)^{a_1-1}
        \Bigg(\frac{\partial}{\partial t_2}\Bigg)^{a_2-1}
        \Bigg(\frac{\partial}{\partial t_3}\Bigg)^{a_3-1}
        GF_{2,\hat{3}}(\vec{t}) \Bigg|_{t_1,t_2,t_3=0} \nonumber\\
        &= H_1(a_1,a_2,a_3) \cdot \sideset{_2}{_1}{\mathop{\mathrm{F}}}\left(\begin{array}{c}
        D-2 ,\frac{D-1}{2}\\D-1
        \end{array} \Bigg | 1-\frac{z^{-}_3}{z^{+}_3}\right)
        + H_0(a_1,a_2,a_3).
      \end{align}
\end{small}
Let us take $a_1=1,a_2=1,a_3=2$ as the example, then we have $H_0(1,1,2)$ is 
\begin{footnotesize}
     \begin{align}
        & \frac{D-2}{m_1^4+(m_2^2-m_3^2)^2-2m_1^2(m_2^2+m_3^2)} 
        \Bigg[ \Bigg(\frac{m_1^2+m_2^2-m_3^2-2\sqrt{m_1^2m_2^2}}{m_1^2+m_2^2-m_3^2+2\sqrt{m_1^2m_2^2}}\Bigg)^{\frac{1-D}{2}}
        \Bigg(\frac{m_1^2+m_2^2-m_3^2+2\sqrt{m_1^2m_2^2}}{m_1^2+m_2^2-m_3^2-2\sqrt{m_1^2m_2^2}}\Bigg)^{\frac{1-D}{2}}\Bigg]\nonumber\\
        &= \frac{D-2}{m_1^4+(m_2^2-m_3^2)^2-2m_1^2(m_2^2+m_3^2)}.
      \end{align}
\end{footnotesize}

    It can be seen that the analytic (irrational) parts of $H_0(1,1,2)$ cancel out, and the remaining expression is purely rational. Moreover, the resulting rational function precisely matches the expected reduction coefficient. One can also check that for $H_1(1,1,2)$, there is
\begin{footnotesize}
    \begin{align}
         \Bigg(\frac{m_1^2+m_2^2-m_3^2+2\sqrt{m_1^2m_2^2}}{m_1^2+m_2^2-m_3^2-2\sqrt{m_1^2m_2^2}}\Bigg)^{\frac{-1-D}{2}}
        \Big[\frac{(m_1^2+m_2^2-m_3^2)(D-3)}{(m_1^2+m_2^2-m_3^2-2\sqrt{m_1^2m_2^2})^3} -
        \frac{(m_1^2+m_2^2-m_3^2)(D-3)}{(m_1^2+m_2^2-m_3^2-2\sqrt{m_1^2m_2^2})^3}    
        \Big],
      \end{align}
\end{footnotesize}
which indeed vanishes.

It is important to note that \textbf{one should perform analytic simplification before substituting numerical values when using generating functions to extract reduction coefficients}, because the expansion is generally defined over the complex domain. If this step is skipped, some analytic cancellations may fail to manifest numerically, potentially leading to spurious complex values—even if the final result is in fact real. For instance, in the case of $H_0(1,1,2)$, the cancellation inside the brackets occurs at the symbolic level. If we substitute numerical values prematurely, such as $m_1 = 2.42$, $m_2 = 3.57$, $m_3 = 5.38$, and $D = 5.28$, we find:
\begin{align}
&\Bigg(\frac{m_1^2 + m_2^2 - m_3^2 - 2\sqrt{m_1^2 m_2^2}}{m_1^2 + m_2^2 - m_3^2 + 2\sqrt{m_1^2 m_2^2}}\Bigg)^{\frac{1-D}{2}} 
\Bigg(\frac{m_1^2 + m_2^2 - m_3^2 + 2\sqrt{m_1^2 m_2^2}}{m_1^2 + m_2^2 - m_3^2 - 2\sqrt{m_1^2 m_2^2}}\Bigg)^{\frac{1-D}{2}} \nonumber \\
&= (-3.98257)^{\frac{1 - 5.28}{2}} \cdot \left(\frac{1}{-3.98257}\right)^{\frac{1 - 5.28}{2}} = 0.637424 - 0.770513 i.
\end{align}
This result arises solely due to evaluating the expression numerically before performing the analytic simplification, which would have led to a real-valued result. Therefore, analytic preprocessing is necessary to ensure the correct interpretation of such expressions.

We have explicitly verified that after applying the above transformation, $ H_1(\vec{a})$ vanishes and the irrational parts in $H_0(\vec{a})$ indeed cancel out through division. Taking advantage of this property, one can safely discard the hypergeometric function terms that appear after differentiation when performing practical calculations—whether implemented in code or carried out manually. Moreover, any remaining elementary terms in $H_0(\vec{a})$ expressions such as $ F(m_i^2, q_i \cdot q_j)^{D/2} $ or $ F(m_i^2, q_i \cdot q_j)^{D} $ can be treated as inert prefactors that cancel in the final rational result. Therefore, in practical implementations, these terms can be effectively replaced by 1 to simplify the computation without affecting the outcome. In addition, for any terms raised to fractional powers (e.g., $1/2$ powers), it is important to first multiply the bases before applying the exponentiation. This ensures that cancellations and simplifications take place correctly. To illustrate the simplification procedure, we now consider a toy example. The expression given below is not derived from an actual computation, but is constructed solely for the purpose of demonstrating how the simplification works in practice.
\begin{align}
    &\Bigg[
    \frac{(D-3)}{m_1^2}  +\dots
    \Bigg] 
    \left(\frac{m_1^2-\sqrt{m_1^2p^2}}{m_1^2-\sqrt{m_1^2p^2}}\right)^{\frac{D-1}{2}} 
    \cdot \sideset{_2}{_1}{\mathop{\mathrm{F}}}\left( \begin{array}{c} \frac{D-1}{2},D-2 \\ D-1 \end{array} \middle| 1-\frac{m_1^2+p^2-2\sqrt{m_1^2p^2}}{m_1^2+p^2+2\sqrt{m_1^2p^2}}  \right) \nonumber\\
    &+ \frac{(D-2)(p^2)^{\frac{D-1}{2}}(p^2)^{-\frac{D+3}{2}}}{m_1^4-2 m_1^2 p^2}
    \left(\frac{m_1^2+p^2-2 \sqrt{m_1^2 p^2}}{m_1^2+p^2+2 \sqrt{m_1^2 p^2}}\right)^{\frac{1-D}{2}} 
    \left(\frac{m_1^2+p^2+2 \sqrt{m_1^2 p^2}}{m_1^2+p^2-2\sqrt{m_1^2 p^2}}\right)^{\frac{1-D}{2}}\nonumber\\
    =&\frac{(D-2)(p^2)^{\frac{D-1}{2}}(p^2)^{-\frac{D+3}{2}}}{m_1^4-2 m_1^2 p^2}
    \left(\frac{m_1^2+p^2-2 \sqrt{m_1^2 p^2}}{m_1^2+p^2+2 \sqrt{m_1^2 p^2}}\right)^{\frac{1-D}{2}} 
    \left(\frac{m_1^2+p^2+2 \sqrt{m_1^2 p^2}}{m_1^2+p^2-2\sqrt{m_1^2 p^2}}\right)^{\frac{1-D}{2}}\nonumber\\
    =&\frac{(D-2)(\frac{1}{p^2})^{\frac{1}{2}}(\frac{1}{p^2})^{\frac{1}{2}}}{(m_1^4-2 m_1^2 p^2)p^2}
    \left(\frac{m_1^2+p^2-2 \sqrt{m_1^2 p^2}}{m_1^2+p^2+2 \sqrt{m_1^2 p^2}}\right)^{\frac{1}{2}} 
    \left(\frac{m_1^2+p^2+2 \sqrt{m_1^2 p^2}}{m_1^2+p^2-2\sqrt{m_1^2 p^2}}\right)^{\frac{1}{2}}\nonumber\\
    =&\frac{(D-2)}{(m_1^4-2m_1^2p^2)p^2}\Bigg[(\frac{1}{p^2})(\frac{1}{p^2})\left(\frac{m_1^2+p^2-2 \sqrt{m_1^2 p^2}}{m_1^2+p^2+2 \sqrt{m_1^2 p^2}}\right) 
    \left(\frac{m_1^2+p^2+2 \sqrt{m_1^2 p^2}}{m_1^2+p^2-2\sqrt{m_1^2 p^2}}\right)\Bigg]^{1/2}\nonumber\\
    =&\frac{(D-2)}{(m_1^4-2m_1^2p^2)p^2}\Bigg[(\frac{1}{p^2})^2\Bigg]^{\frac{1}{2}}\nonumber\\
    =&\frac{(D-2)}{(m_1^4-2m_1^2p^2)p^4}.
\end{align}
In the above procedure, the simplification consists of three steps. First, we directly discard the hypergeometric function terms without the need to expand or verify their cancellation explicitly, as we have already established that they vanish in the final rational result. Second, all elementary prefactors of the form \( F(m_i^2, q_i \cdot q_j)^{D/2} \) or \( F(m_i^2, q_i \cdot q_j)^{D} \) can also be replaced by 1 without detailed simplification, since they ultimately cancel as well. At this stage, we are left only with terms raised to the one-half power. In the third step, for all such square-root-like terms, we first multiply the bases together before taking the square root. This straightforward and rule-based approach leads to a fully simplified expression and is particularly helpful for both analytical manipulation and numerical implementation.

\subsection{Type III}\label{sec: 7.3}

The third type of generating function involves hypergeometric functions and arises in cases where one of the integration limits satisfies $z_i^+ = 0$ and $z_i^-$ is rational. Thus, the transformation of above subsection is no longer valid. Examples of this type can be found in Section \ref{sec: 6.2} and in the diagram shown in figure \ref{fig:other example} (a) of Section \ref{sec: 6.4}. A key feature of these Feynman integrals is that the maximum topology does not correspond to a master integral. This behavior can be naturally understood in the Baikov representation: when $z_i^+ = 0$, the Baikov polynomial satisfies $P(\vec{0}) = 0$, which implies that the integral corresponding to the maximum topology vanishes upon taking the residue at $z_i = 0$ for all $i$. As a result, the maximum topology does not contribute as a master integral in this representation. This type of generating functions has the following form:
\begin{align}
  GF(\vec{t}) &= 
     \left(\frac{C'(\vec{t})}{C}\right)^\gamma
     \Bigg(\frac{(\tilde{z}^+(\vec{t}) - \tilde{z}^-(\vec{t}))^2}{(z^+-z^-)^2} \Bigg)^{\frac{2\gamma+1}{2}}
     \frac{1}{\tilde{z}^-\vec{t})}
     \sideset{_2}{_1}{\mathop{\mathrm{F}}}\left( \begin{array}{c} 1,\gamma+1\\ 2\gamma+2 \end{array} \middle| 1-\frac{\tilde{z}^+(\vec{t})}{\tilde{z}^-(\vec{t})}  \right) .
     \label{eq:type3GF}
\end{align}

The derivative formula for the hypergeometric function is
\begin{equation}
    \frac{d}{dz}  \sideset{_2}{_1}{\mathop{\mathrm{F}}}\left(\begin{array}{c}
    a ,b\\  c
    \end{array} \Bigg | z\right)
    = \frac{ab}{c} \sideset{_2}{_1}{\mathop{\mathrm{F}}}\left(\begin{array}{c}
    1+a,1+b\\  1+c
    \end{array} \Bigg | z\right) .
    \label{eq:diff}
\end{equation}
Moreover, at $\vec{t} = \vec{0}$, we have $C'(\vec{t}) = C$, $\tilde{z}_i^-(\vec{0}) = z_i^-$, $\tilde{z}_i^+(\vec{0}) = z_i^+=0$ and $z^-(\vec{0})$ is also rational. Therefore, after differentiation, the only potentially irrational term in the generating function is the hypergeometric function with form:
\begin{equation}
    \sideset{_2}{_1}{\mathop{\mathrm{F}}}\left(\begin{array}{c}
    n+1,\gamma+1+n\\  2\gamma+2+n
    \end{array} \Bigg | 1\right) ,
    \label{eq:2F1eq}
\end{equation}
where $n$ is an integer. The variable in the hypergeometric function is one. We can use the formula:
\begin{equation}
  \sideset{_2}{_1}{\mathop{\mathrm{F}}}\left( \begin{array}{c} a,b\\c \end{array} \middle| 1 \right)
  = \frac{\Gamma(c)\Gamma(c-a-b)}{\Gamma(c-a)\Gamma(c-b)},
  \label{eq:trans 3-1}
\end{equation}
to transform it into rational form as
\begin{equation}
    \sideset{_2}{_1}{\mathop{\mathrm{F}}}\left(\begin{array}{c}
    n+1,\gamma+1+n\\  2\gamma+2+n
    \end{array} \Bigg | 1\right)=\frac{\Gamma(2\gamma+2+n)\Gamma(\gamma-n)}{\Gamma(\gamma+1)\Gamma(2\gamma+n+1)}=\frac{\left(2\gamma+n+1\right)_{n+1}}{\left(\gamma-n\right)_{n+1}} .
\end{equation}



For the type-III case with form \eqref{eq:type3GF}, one point should be emphasized: complete asymptotic expansion yields non-integer powers, but not having any impact on extracting reduction coefficients. For example the expansion of \eqref{eq:GF of Suncut} is:

\begin{equation}
    GF_{4, \hat{5}}(\vec{t})=\frac{\pi \Gamma(D-2)}{s \sin{(\frac{\pi D}{2})}\Gamma(\frac{D}{2}-1)^2} \left(\frac{t_5}{s}\right)^{D/2-2}-\frac{2(D-3)}{(D-4)s}+...
\end{equation}

The term with non-integer power is expected from the expansion-by-regions method. However, the reduction coefficients derived by \eqref{eq:nton} are only related to the limit of the derivations of the generating function at $t\rightarrow 0$. Arbitrary $n$ order derivation of the term with non-integer power is:
\begin{equation}
    \frac{d^n}{dt_5^n} \left(\frac{t_5}{s}\right)^{D/2-2} = \frac{\Gamma(\frac{D}{2}-1)}{\Gamma(\frac{D}{2}-1-n)} \left(\frac{t_5}{s}\right)^{D/2-2-n}.
\end{equation}
Due to the arbitrariness of dimensional regulater $D$, for arbitrary $n$ we have\footnote{In fact, \eqref{eq:2F1eq} strictly holds under the condition $c - a - b > 0$. Thus, when we take \red{higher-order} terms in the parameters $t_i$, we assume the spacetime dimension $D$ is sufficiently large so that the non-integer powers do not affect the low-order coefficients in $t_i$.}
\begin{equation}
    t_5^{D/2-2-n}|_{t\rightarrow 0} = 0.
\end{equation}
Therefore, although the term with non-integer power is \red{in} confict with the definition of generating function, it is not related to the reduction coefficients derived from generating function, and it's safe to use \eqref{eq:type3GF} as the formal summation of series.

\section{Summary}\label{sec: 8}

In this paper, we revisit the residue method in the Baikov representation and use it to study generating functions for reduction coefficients. Compared to the methods and results in \cite{Feng:2024qsa,Li:2024rvo} , this paper either provides a more concise expression for the generating functions or achieves expressions of similar simplicity through a less complex process. In \cite{Li:2024rvo}, $GF_{N-1;\hat{j}}(\vec{t})$ is obtained through a simple single-variable integral. However, the result is expressed as an infinite series, where each term involves an Appell function. In \cite{Feng:2024qsa}, a concise expression containing only two hypergeometric function is provided. However, deriving this result requires establishing a system of multivariable partial differential equations, analyzing the relationships satisfied by their coefficients, and utilizing several lemmas to rigorously prove the final form. In contrast, the method in this paper requires only a single definite integral to produce an expression with just two hypergeometric functions.  By this method, we extend the previous results to one-loop integrals with non-standard quadratic propagators and some multi-loop cases. 
This means that, compared with the momentum-space representation, the Baikov representation makes common features and analytic structure of Feynman integrals easier to see. In addition, when some closed-form generating functions are expanded naively, non-integer powers appear; however, this does not affect how we extract the reduction coefficients. It suggests that allowing such non-integer terms may lead to a more compact, modified generating function. A further advantage of generating function is that it is often simpler than the coefficients themselves. For example, it takes hypergeometric function forms. So we can use the many known identities and differential relations of hypergeometric functions to study the reduction coefficients.

The limitations of this work are clear: our results are still restricted to the top and sub top sectors. Our study of reductions to the lower sectors is still not in depth, and the conclusions are limited. Besides, our higher-loop examples are relatively simple. These are precisely the directions we plan to study next. For lower-topology reductions, we may start from  \cite{Baikov:1996rk}. For more general situations, we will try other representations or reduction methods—such as the Feynman parametrization, the fixed branch integral (FBI) representation \cite{Huang:2024nij}, and the intersection-number \cite{Mastrolia:2018uzb, Frellesvig:2019kgj,  Frellesvig:2020qot,Weinzierl:2020xyy} approach—together with generating functions, to see whether more conclusion or clearer analytic structure can be obtained.

\section*{Acknowledgments}
We would like to thank Prof. Bo Feng and Prof. Yang Zhang for their valuable advice. We are also grateful to Dr. Xu-Hang Jiang and Dr. Wen Chen for their helpful discussions. 
 As the authors do not have English as their native language, we have used Large Language Models (LLM) to help us polish the article during the writing process.
Wen-Di Li acknowledges the financial support from the Hangzhou Institute for Advanced Study. Xiang Li acknowledges the support from Peking University. Chang Hu expresses gratitude to Hebei University for providing the startup fund for young faculty research.

\renewcommand{\thefootnote}{\arabic{footnote}}
\setcounter{footnote}{0} 
	\flushbottom
	
\appendix
\section{Analytical and numerical results for the reduction coefficients}\label{sec: appendix}
In this appendix, we present the results of several types of integrals to demonstrate the validity of the generating function method used in this paper. All results are compared with those obtained using the FIRE package \cite{Smirnov:2023yhb}, and we will see that the results from both methods are exactly equal.

\subsection{The sunset-type diagram with four propagators}\label{sec: appendix 1}
For the Type-I integrals discussed in Section \ref{sec: 7.1}, we consider the sunset-type diagram with four propagators (figure \ref{Fig:suncut with 4 prop}) $I(a_1,a_2,a_3,a_4)$ as an example. The generating function for this integral, which reduces it to the master integral of the top sector, $I(1,1,1,1)$, is given by \eqref{eq:GF of suncut with 4 prop}. The generating function is an elementary function, and its differentiation is straightforward. By differentiating \eqref{eq:GF of suncut with 4 prop}, we obtain the reduction coefficients for the corresponding integrals, which are listed in table \ref{table:suncut with 4 prop} .
\begin{table}[h]
    \tiny
    \centering
    \begin{tabular}{c|c|c}
    \hline
    $(\vec{a})$ & Method & Analytical expression for the reduction coefficient \\ \hline
    \multirow{2}{*}{(1,1,2,2)} & FIRE  & $\frac{(D-3)^2}{12m^4-3m^2p^2}$ \\ \cline{2-3}
    & GF & $\frac{(D-3)^2}{12m^4-3m^2p^2}$ \\     
    \hline
    \multirow{2}{*}{(2,2,2,2)} & FIRE  & $\frac{(D-3)^2\left(24(D-2)m^4+2(D-14)(D+2)m^2p^2+((D-6)D+32)p^4\right)}{54m^6p^2\left(p^2-4m^2\right)^2}$ \\ \cline{2-3}
    & GF & $\frac{(D-3)^2\left(24(D-2)m^4+2(D-14)(D+2)m^2p^2+((D-6)D+32)p^4\right)}{54m^6p^2\left(p^2-4m^2\right)^2}$ \\     
    \hline
    \multirow{2}{*}{(1,1,2,5)} & FIRE  & $\frac{(D-5)(D-3)^2((D-34)D+216)}{1944m^8\left(4m^2-p^2\right)}$ \\ \cline{2-3}
    & GF & $\frac{(D-5)(D-3)^2((D-34)D+216)}{1944m^8\left(4m^2-p^2\right)}$ \\     
    \hline
    \multirow{2}{*}{(4,1,3,2)} & FIRE  & $\frac{(D-5)(D-3)^2((D-16)D+36)\left((D-4)p^2-4m^2\right)}{972m^8p^2\left(p^2-4m^2\right)^2}$ \\ \cline{2-3}
    & GF & $\frac{(D-5)(D-3)^2((D-16)D+36)\left((D-4)p^2-4m^2\right)}{972m^8p^2\left(p^2-4m^2\right)^2}$ \\     
    \hline
    \multirow{2}{*}{(3,2,3,2)} & FIRE  & $\frac{(D-5)(D-3)^2\left(8(D(5D-47)+156)m^4+2(D((D-25)D+120)-192)m^2p^2+(D-4)((D-10)D+48)p^4\right)}{648m^8p^2\left(4m^2-p^2\right)^3}$ \\ \cline{2-3}
    & GF & $\frac{(D-5)(D-3)^2\left(8(D(5D-47)+156)m^4+2(D((D-25)D+120)-192)m^2p^2+(D-4)((D-10)D+48)p^4\right)}{648m^8p^2\left(4m^2-p^2\right)^3}$ \\     
    \hline
    \multirow{2}{*}{(1,1,8,1)} & FIRE  & $\frac{(D-9)(D-7)(D-5)(D-3)\left(1680(D-8)m^4p^2-84(D-8)(D-6)m^2p^4+(D-8)(D-6)(D-4)p^6-6720m^6\right)}{5040p^6\left(4m^2-p^2\right)^7}$ \\ \cline{2-3}
    & GF & $\frac{(D-9)(D-7)(D-5)(D-3)\left(1680(D-8)m^4p^2-84(D-8)(D-6)m^2p^4+(D-8)(D-6)(D-4)p^6-6720m^6\right)}{5040p^6\left(4m^2-p^2\right)^7}$ \\     
    \hline
    \end{tabular}
    \caption{This table presents the reduction coefficients of the sunset-type diagram with four propagators (figure \ref{Fig:suncut with 4 prop}) $I(a_1,a_2,a_3,a_4)$ to the top sector of the master integrals $I(1,1,1,1)$. We compare the results obtained using the Generating Function Method (denoted as GF) and the FIRE results. }
    \label{table:suncut with 4 prop}
\end{table}

\newpage 
\subsection{Two loop and three loop vacuum diagram}\label{sec: appendix 2}
For the Type-II integrals discussed in section \ref{sec: 7.2}, we consider two loop and three loop vacuum diagram (figure \ref{fig:23loopvac}) $I_{2-vac}(a_1,a_2,a_3), I_{3-vac}(a_1,a_2,a_3,a_4,a_5,a_6)$ as examples. 
The generating functions for reducing these two integrals to their respective sub \red{top sector} master integrals can both be given by \eqref{eq:GF of type II}.
By differentiating the generating functions, we can obtain the reduction coefficients. In table \ref{table:2 loop vacuum}, we provide the analytical expressions for the reduction coefficients of the two-loop vacuum diagrams. 
Since the expressions for the three-loop vacuum diagrams are too lengthy, we present the numerical results for the three-loop vacuum diagrams in table \ref{table:3 loop vacuum}. 
Moreover, we employ the techniques described in section \ref{sec: 7.2}, where we find that the coefficient of the hypergeometric function equals zero, and we directly set the terms with powers of the dimension \( D \) to be equal to 1.
Feynman
\begin{table}[h]
    \tiny
    \centering
    \begin{tabular}{c|c|c}
    \hline
    $(\vec{a})$ & Method & Analytical expression for the reduction coefficient \\ \hline
    \multirow{2}{*}{(1,1,2)} & FIRE  & $\frac{D-2}{m_1^4-2\left(m_2^2+m_3^2\right)m_1^2+\left(m_2^2-m_3^2\right){}^2}$ \\ \cline{2-3}
    & GF & $\frac{D-2}{m_1^4-2\left(m_2^2+m_3^2\right)m_1^2+\left(m_2^2-m_3^2\right){}^2}$ \\     
    \hline
    \multirow{2}{*}{(1,1,3)} 
    & FIRE  & $-\frac{(D-5)(D-2)\left(m_1^2+m_2^2-m_3^2\right)}{2\left(m_1^4-2\left(m_2^2+m_3^2\right)m_1^2+\left(m_2^2-m_3^2\right){}^2\right){}^2}$ \\ \cline{2-3}
    & GF & $-\frac{(D-5)(D-2)\left(m_1^2+m_2^2-m_3^2\right)}{2\left(m_1^4-2\left(m_2^2+m_3^2\right)m_1^2+\left(m_2^2-m_3^2\right){}^2\right){}^2}$ \\     
    \hline
    \multirow{2}{*}{(1,1,4)} 
    & FIRE  & $\frac{(D-2)\left(((D-10)D+27)m_1^4+2m_1^2\left(((D-14)D+43)m_2^2-((D-10)D+27)m_3^2\right)+((D-10)D+27)\left(m_2^2-m_3^2\right){}^2\right)}{6\left(m_1^4-2\left(m_2^2+m_3^2\right)m_1^2+\left(m_2^2-m_3^2\right){}^2\right){}^3}$ \\ \cline{2-3}
    & GF & $\frac{(D-2)\left(((D-10)D+27)m_1^4+2m_1^2\left(((D-14)D+43)m_2^2-((D-10)D+27)m_3^2\right)+((D-10)D+27)\left(m_2^2-m_3^2\right){}^2\right)}{6\left(m_1^4-2\left(m_2^2+m_3^2\right)m_1^2+\left(m_2^2-m_3^2\right){}^2\right){}^3}$ \\     
    \hline
    \multirow{2}{*}{(3,1,1)} & FIRE  & $\frac{(D-2)\left(-3(D-4)m_1^6+m_1^4\left(7(D-4)m_2^2+Dm_3^2\right)-\left(m_2^2-m_3^2\right)m_1^2\left(5(D-4)m_2^2+(3D-16)m_3^2\right)+(D-4)\left(m_2^2-m_3^2\right){}^3\right)}{8m_1^4\left(m_1^4-2\left(m_2^2+m_3^2\right)m_1^2+\left(m_2^2-m_3^2\right){}^2\right){}^2}$  \\ \cline{2-3}
    & GF & $\frac{(D-2)\left(-3(D-4)m_1^6+m_1^4\left(7(D-4)m_2^2+Dm_3^2\right)-\left(m_2^2-m_3^2\right)m_1^2\left(5(D-4)m_2^2+(3D-16)m_3^2\right)+(D-4)\left(m_2^2-m_3^2\right){}^3\right)}{8m_1^4\left(m_1^4-2\left(m_2^2+m_3^2\right)m_1^2+\left(m_2^2-m_3^2\right){}^2\right){}^2}$\\     
    \hline
    \multirow{2}{*}{(1,2,2)} & FIRE  & $\frac{(D-2)\left(-2m_1^2\left((D-4)m_2^2+m_3^2\right)+(2D-9)m_2^4-2(D-4)m_2^2m_3^2+m_1^4+m_3^4\right)}{2m_2^2\left(m_1^4-2\left(m_2^2+m_3^2\right)m_1^2+\left(m_2^2-m_3^2\right){}^2\right){}^2}$  \\ \cline{2-3}
    & GF & $\frac{(D-2)\left(-2m_1^2\left((D-4)m_2^2+m_3^2\right)+(2D-9)m_2^4-2(D-4)m_2^2m_3^2+m_1^4+m_3^4\right)}{2m_2^2\left(m_1^4-2\left(m_2^2+m_3^2\right)m_1^2+\left(m_2^2-m_3^2\right){}^2\right){}^2}$  \\     
    \hline
    \end{tabular}
    \caption{This table presents the reduction coefficients of the two loop vacuum diagram, (figure \ref{fig:23loopvac}(a)) $I(a_1,a_2,a_3)$ to the sub top sector of the master integrals $I(1,1,0)$. We compare the results obtained using the Generating Function Method (denoted as GF) and the FIRE results. Where we found that the coefficient \( H_1(\vec{a}) \) of the hypergeometric function equals zero, and $H_0(\vec{a})$ is agree with FIRE.}
    \label{table:2 loop vacuum}
\end{table}

\begin{table}[h] 
    
    \centering 
    \begin{tabular}{c|c|c|c} 
    \hline 
     $(\vec{a})$  &  FIRE 
     &  $GF\big(H_0(\vec{a})\big)$ & $GF\big(H_1(\vec{a})\big)$ \\ \hline  
    (1,1,1,1,1,8) & $2.72033\times10^{-6}$ & $2.72033\times10^{-6}$ & 0 \\ \hline
    (1,1,1,1,2,2) & 0.00874267 & 0.00874267 & 0  \\ \hline
    (1,1,2,2,2,2) & 0.0217022 & 0.0217022 & 0 \\ \hline
    (2,2,2,2,2,2) & -0.00747571 & -0.00747571 & 0 \\ \hline
    (1,5,1,3,1,2) & 0.000425422 & 0.000425422 & 0 \\ \hline
    (6,1,3,2,2,1) & $-6.57283\times 10^{-6}$ & $-6.57283\times 10^{-6}$ &0 \\ \hline
    \end{tabular}
    \caption{ This table presents the numerical results for the reduction coefficients of the three loop vacuum diagram, (figure \ref{fig:23loopvac}(b)) $I(a_1,a_2,a_3,a_4,a_5,a_6)$ to the sub top sector of the master integrals $I(0,1,1,1,1,1)$. We compare the results obtained using the Generating Function Method (denoted as GF) and the FIRE results. We set $m^2=2.45,D=8.23$. Where we found that the coefficient \( H_1(\vec{a}) \) of the hypergeometric function equals zero, and $H_0(\vec{a})$ is agree with FIRE.}.  
    \label{table:3 loop vacuum}
\end{table}

\newpage
\subsection{The massless sunset-type diagram with a vertical propagator}\label{sec: appendix 3}
For the Type-III integrals discussed in section \ref{sec: 7.3}, we consider the massless sunset-type diagram with a vertical propagator (figure \ref{fig:suncut 01}) $I(a_1,a_2,a_3,a_4,a_5)$ as an example. The generating function for this integral, which reduces it to the master integral of the sub top sector $I(1,1,1,0)$, is given by \eqref{eq:GF of Suncut}. By differentiating \eqref{eq:GF of Suncut}, we obtain the reduction coefficients for the corresponding integrals, which are listed in table \ref{table:suncut}.
\begin{table}[h]
    \small
    \centering
    \begin{tabular}{c|c|c}
    \hline
    $(\vec{a})$ & Method & Analytical expression for the reduction coefficient \\ \hline
    \multirow{2}{*}{(2,3,3,3,3)} & FIRE  & $\frac{(D-9)(D-8)(D-7)(D-5)(D-3)(D(D(D((D-61)D+1482)-17888)+107024)-253536)}{(D-14)(D-12)(D-10)(p^2)^{10}}$ \\ \cline{2-3}
    & GF & $\frac{(D-9)(D-8)(D-7)(D-5)(D-3)(D(D(D((D-61)D+1482)-17888)+107024)-253536)}{(D-14)(D-12)(D-10)(p^2)^{10}}$ \\     
    \hline
    \multirow{2}{*}{(1,1,1,1,6)} & FIRE  & $\frac{64(D-7)(D-5)(D-3)}{(D-14)(D-12)(D-10)(p^2)^6}$ \\ \cline{2-3}
    & GF & $\frac{64(D-7)(D-5)(D-3)}{(D-14)(D-12)(D-10)(p^2)^6}$ \\     
    \hline
    \multirow{2}{*}{(1,1,1,10,1)} & FIRE  & $\frac{(D-12)(D-11)(D-10)(D-9)(D-8)(D-7)(D-6)(D-5)(D-3)}{181440(p^2)^{10}}$ \\ \cline{2-3}
    & GF & $\frac{(D-12)(D-11)(D-10)(D-9)(D-8)(D-7)(D-6)(D-5)(D-3)}{181440(p^2)^{10}}$ \\     
    \hline
    \multirow{2}{*}{(1,12,1,1,1)} & FIRE  & $\frac{(D-14)(D-13)(D-12)(D-11)(D-10)(D-9)(D-8)(D-7)(D-6)(D-5)(D-3)}{19958400(p^2)^{12}}$ \\ \cline{2-3}
    & GF & $\frac{(D-14)(D-13)(D-12)(D-11)(D-10)(D-9)(D-8)(D-7)(D-6)(D-5)(D-3)}{19958400(p^2)^{12}}$ \\     
    \hline
    \multirow{2}{*}{(1,1,1,5,5)} & FIRE  & $-\frac{4(D-11)(D-9)(D-7)(D-5)(D-3)}{3(D-12)(p^2)^9}$ \\ \cline{2-3}
    & GF & $-\frac{4(D-11)(D-9)(D-7)(D-5)(D-3)}{3(D-12)(p^2)^9}$ \\     
    \hline
    \multirow{2}{*}{(1,1,1,5,8)} & FIRE  & $\frac{32(D-13)(D-11)(D-9)(D-7)(D-5)(D-3)}{3(D-18)(D-16)(p^2)^{12}}$ \\ \cline{2-3}
    & GF & $\frac{32(D-13)(D-11)(D-9)(D-7)(D-5)(D-3)}{3(D-18)(D-16)(p^2)^{12}}$ \\     
    \hline
    \end{tabular}
    \caption{This table presents the reduction coefficients of the massless sunset-type diagram with a vertical propagator (figure \ref{fig:suncut 01}) $I(a_1,a_2,a_3,a_4,a_5)$ to the sub top sector of the master integrals $I(1,1,1,1,0)$. We compare the results obtained using the Generating Function Method (denoted as GF) and the FIRE results.}
    \label{table:suncut}
\end{table}

\newpage
	
\bibliographystyle{JHEP}
\bibliography{reference}

\end{document}